\documentclass[journal, draftcls, onecolumn, 12pt]{IEEEtran}

\usepackage[T1]{fontenc}
\usepackage{aeguill}
\usepackage{mathdots}
\usepackage{amsmath}%,amsthm}
\usepackage{amssymb}
\usepackage[all]{xy}
\usepackage{color}
\usepackage{lscape}
\usepackage{graphics}
\usepackage{psfrag}
\usepackage{graphicx}
\usepackage{theorem}

\renewcommand {\epsilon} {\varepsilon}

\newcommand{\abs}[1]{\left\lvert#1\right\rvert}
\newcommand{\nrm}[1]{\left\lVert#1\right\rVert}

\newcommand{\vect}[1]{\boldsymbol{#1}}

\newcommand{\tendvers}[1]{\xrightarrow[n\rightarrow\infty]{}#1}

\newcommand{\Sinf}{\mathcal{S}_\infty}
\newcommand{\Linf}{\mathcal{L}_\infty}

\def\N{{\mathbb N}}        % integers
\def\R{{\mathbb R}}        % reals
\def\C{{\mathbb C}}        % complexs
\def\P{{\mathbb P}}        % probability
\def\E{{\mathbb E}}        % expectation
\def\1{{\mathbf 1}}        % indicator
\def\ep{\varepsilon}

\newtheorem{thm}[equation]{Theorem}
\newtheorem{prop}[equation]{Proposition}
\newtheorem{cor}[equation]{Corollary}

\newtheorem{lem}[equation]{Lemma}

\newtheorem{rmk}[equation]{Remark}

\allowdisplaybreaks

\DeclareMathOperator{\Id}{Id}

\DeclareMathOperator{\vct}{vect}

\newlength{\minipagewidth}
\title{On Information Rates of the Fading Wyner Cellular Model via the Thouless Formula for the Strip}

\author{\authorblockN{Nathan Levy\authorrefmark{1}\authorrefmark{2}, Ofer Zeitouni\authorrefmark{3}\authorrefmark{4} and Shlomo Shamai (Shitz)\authorrefmark{2}}\\
\authorblockA{\authorrefmark{1}
D\'epartement de Math\'ematiques et Applications, Ecole Normale Sup\'erieure, Paris 75005,
France}\\
\authorblockA{\authorrefmark{2}
Department of Electrical Engineering, Technion, Haifa 32000,
Israel}\\
\authorblockA{\authorrefmark{3}
Department of Mathematics, Weizmann Institute of Science, Rehovot 76100, Israel}\\
\authorblockA{\authorrefmark{4}
School of Mathematics, University of Minnesota, Minneapolis, MN 55455,
USA}\\
}
\date{\today}

\begin{document}
\maketitle

\begin{abstract}
We apply the theory of random Schr\"odinger operators to the analysis of multi-users communication channels similar to the Wyner model, that are characterized by short-range intra-cell broadcasting. With $H$ the channel transfer matrix, $HH^\dagger$ is a narrow-band matrix and in many aspects is similar to a random Schr\"odinger operator. We relate the per-cell sum-rate capacity of the channel to the integrated density of states of a random Schr\"odinger operator; the latter is related to the top Lyapunov exponent of a random sequence of matrices via a version of the Thouless formula. Unlike related results in classical random matrix theory, limiting results do depend on the underlying fading distributions. We also derive several bounds on the limiting per-cell sum-rate capacity, some based on the theory of random Schr\"odinger operators, and some derived from information theoretical considerations. Finally, we get explicit results in the high-SNR regime for some particular cases.
\end{abstract}

%%%%%%%%%%%%%%%%%%%%%%%%%%%%%%%%%%%%%%%%%
%%%%%%%%%%%%%%%%%%%%%%%%%%%%%%%%%%%%%%%%%
%%%%%%%%%%%%%%%%%%%%%%%%%%%%%%%%%%%%%%%%%

\section{Introduction}
\label{sec:Introduction}

%%%%%%%%%%%%%%%%%%%%%%%%%%%%%%%%%%%%%%%%%
%%%%%%%%%%%%%%%%%%%%%%%%%%%%%%%%%%%%%%%%%
%%%%%%%%%%%%%%%%%%%%%%%%%%%%%%%%%%%%%%%%%

The growing demand for ubiquitous access to high-data rate services, has produced a huge amount of research analyzing the performance of wireless communications systems. Techniques for providing better service and coverage in cellular mobile communications are currently being investigated by industry and academia. In particular, the use of joint multi-cell processing (MCP), which allows the base-stations (BSts) to jointly process their signals, equivalently creating a distributed antenna array, has been identified as a key tool for enhancing system performance (see
\cite{Somekh-Simeone-Barness-Haimovich-Shamai-BookChapt-07}\cite{Shamai-Somekh-Zaidel-JWCC-2004} and references therein for surveys of recent results on multi-cell processing).

Motivated by the fact that mobile users in a cellular system ``see'' only a small number of BSts, and by the desire to provide analytical results, an attractive analytically tractable model for a multi-cell system was suggested by Wyner in \cite{Wyner-94} (see also \cite{Hanly-Whiting-Telc-1993} for an earlier relevant work). In this model, the system's cells are ordered in either an infinite linear array, or in the familiar two-dimensional hexagonal pattern (also infinite). It is assumed that only adjacent-cell interference is present and characterized by a single parameter, a scaling factor $\alpha\in[0,1]$. Considering non-fading channels and a ``wideband'' (WB) transmission scheme, where all bandwidth is available for coding (as opposed to random spreading), the throughputs obtained with optimum and linear MMSE joint processing of the received signals from all cell-sites are derived in \cite{Wyner-94}. Since it was first presented, ``Wyner-like'' models have provided a framework for many works analyzing various transmission schemes in both the up-link and down-link channels (see \cite{Somekh-Simeone-Barness-Haimovich-Shamai-BookChapt-07}\cite{bidiagonal} and references therein).

In this paper we consider a generalized ``Wyner-like" cellular setup and study its per-cell sum-rate capacity. According to Wyner's setup, the cells are arranged on a circle (or a line), and the mobile users ``see" only a fixed number of BSts which are located close to their cell's boundaries. All the BSts are assumed to be connected through an ideal back-haul network to a central multi-cell processor (MCP), that can \emph{jointly} process the up-link received signals of all cell-sites, as well as pre-process the signals to be transmitted by all cell-sites in the down-link channel. The model is characterized by short-range intra-cell broadcasting. Thus, if we denote by $H$ the channel transfer matrix, then $HH^\dagger$ is in many aspects similar to a random Schr\"odinger operator. More specifically, the per-cell sum-rate capacity of the channel is a function of the integrated density of state of $HH^\dagger$, which in turn is related to the top Lyapunov exponent of a random sequence of matrices via a version of the Thouless formula. Unlike associated results in classical random matrix theory, limiting results do depend on the underlying fading distributions.

As an application of our result and motivated by the fact that future cellular systems implicitly assume high-SNR configurations mandatory for high data rate services, we get explicit results in the high-SNR regime for some particular cases.

The rest of the paper is organized as follows. In Section \ref{sec:ProblemStatement}, we present the problem statement. In Section \ref{sec:MainResult}, we prove the convergence of the per-cell sum-rate capacity when the number of cells and BSts goes to infinity and we express the limit in terms of the Lyapunov exponent of a sequence of random matrices (Theorem \ref{principalK}). In Section \ref{sec:Reformulations}, we give several reformulations of this result that yields a particularly simple expression in the high-SNR regime. In Section \ref{sec:BoundsOnTheCapacity}, we give different bounds on the per-cell sum-rate capacity, some of which are based on the theory of product of random matrices, and some on information theoretical considerations. In Section \ref{sec:ResultsForParticularCasesInTheHighSNRRegime}, we specialize the results and make them explicit in some particular cases. Finally in Section \ref{sec:NumericalSimulations} we discuss some open problems using numerical simulations. The relevant background on the theory of Lyapunov exponents is given in Appendix \ref{sec:LyapunovExponentsTheory}, and the relevant background on exterior products is given in Appendix \ref{sec:ExteriorProduct}. Several proofs are postponed to Appendices \ref{sec:ProofOfTheoremRefPrincipalK1}, \ref{sec:ProofOfTheoremRefPrincipalK2} and \ref{sec:OtherProofs}. The per-cell sum-rate capacity of the non-fading channels is derived in Appendix \ref{sec:CapacityOfTheNonFadingChannels}.

%%%%%%%%%%%%%%%%%%%%%%%%%%%%%%%%%%%%%%%%%
%%%%%%%%%%%%%%%%%%%%%%%%%%%%%%%%%%%%%%%%%
%%%%%%%%%%%%%%%%%%%%%%%%%%%%%%%%%%%%%%%%%

\section{Problem statement}
\label{sec:ProblemStatement}

%%%%%%%%%%%%%%%%%%%%%%%%%%%%%%%%%%%%%%%%%
%%%%%%%%%%%%%%%%%%%%%%%%%%%%%%%%%%%%%%%%%
%%%%%%%%%%%%%%%%%%%%%%%%%%%%%%%%%%%%%%%%%

In this paper we consider the following setup. $m+d$ cells with $K$ single antenna users per cell are arranged on a line, where the $m$ single antenna BSts are located in the cells. Starting with the WB transmission scheme where all bandwidth is devoted for coding and all $K$ users are transmitting simultaneously each with average power $\rho$, and assuming synchronized communication, a vector baseband representation of the signals received at the system's BSts is given for an arbitrary time index $i$ by
\[y(i)=H_m(i) x(i)+n(i),\]
where $x(i)$ is the $(m+d)K$ complex Gaussian symbols vector, $z(i)$ is the unitary complex Gaussian additive noise vector. Note that the SNR is $\rho$. From now on, we omit the time index $i$. $H_m$ is the following $m\times K(m+d)$ channel transfer matrix, which is a $d+1$ block diagonal matrix defined by
\[H_m=
\begin{pmatrix}
\zeta_{1,1}& \zeta_{1,2} & \cdots &\zeta_{1,d+1}&     0     &   \cdots  &       0      \\
   0     & \zeta_{2,2} & \cdots &\zeta_{2,d+1}&\zeta_{2,d+2}&           &    \vdots   \\
 \vdots  &           & \ddots &   \ddots  &           &  \ddots   &     0       \\
     0   & \cdots    &    0   & \zeta_{m,m} &\zeta_{m,m+1}&  \cdots   &  	\zeta_{m,d+m}  \\
\end{pmatrix},\]
where $\zeta_{i,j}$ are $1\times K$ row vectors. For $s\in\N^*$, we will denote by $\zeta^s$ the vector $(\zeta_{s-d,s},\dots,\zeta_{s,s})$ and we denote by $\pi$ it distribution. We assume in the rest of the paper that for $n\in\N^*$ and $0\leq i\leq d$ the vectors $(\zeta_{n-i,n})$ are distributed according to $\pi_i$. We define $\Omega=(\zeta^n)_{n\in\N^*}$ and $\P$, the probability distribution on $\Omega$ associated to the above problem. We denote by $\E$ the associated expectation. We also use the 2 norm for vectors and matrices. For matrices, it is the Froebenius norm, which is a sub-multiplicative norm.

\renewcommand{\labelenumi}{(H\arabic{enumi})}

Throughout this paper, we assume a subset of the following hypotheses.
\begin{enumerate}
	\item\label{ergodicity} The vectors $(\zeta^j)_{j\in\N^*}$ form a stationary ergodic sequence.
	\item\label{moments} There exists $\ep>0$ such that for $0\leq i\leq d$, $\E_{\pi_i}\abs{\log\abs{x}}^{1+\ep}<\infty$.
	\item\label{frontier} If $(x_0,\dots,x_d)$ is distributed according to $\pi$, then almost surely, $x_0 x_d^\dagger\neq0$.
	\end{enumerate}

\renewcommand{\labelenumi}{\arabic{enumi}.}

For $m\in\N^*$ and $\lambda>0$, we set $G_m=H_mH_m^\dagger+\lambda\Id_m$, where $\Id_m$ is the $m\times m$ identity matrix. Although $G_m$ depends on $\lambda$, we will not write that dependence unless there is an ambiguity. Under the assumption that $H_m(i)$ is ergodic with respect to the time index $i$, that the Channel State Information (CSI) is known at the receiver whereas the users know the statistics of the CSI, and that the channel varies fast enough so as to allow each transmitted codeword to experience a large number of fading states, we follow \cite{Somekh-Simeone-Barness-Haimovich-Shamai-BookChapt-07} and study the per-cell sum-rate capacity that is given by the following formula (\cite{Telatar-ETT-99}) 
\begin{equation}
\label{capacity}
Cap_{m}(\rho)=\frac{1}{m}\E\log\det\left(\Id+\rho H_m H_m^\dagger\right)=\log\rho+\frac{1}{m}\E\left(\log\det G_m(\lambda)\right),
\end{equation}
where $\lambda=1/\rho$.

%%%%%%%%%%%%%%%%%%%%%%%%%%%%%%%%%%%%%%%%%
%%%%%%%%%%%%%%%%%%%%%%%%%%%%%%%%%%%%%%%%%
%%%%%%%%%%%%%%%%%%%%%%%%%%%%%%%%%%%%%%%%%

\section{Main result}
\label{sec:MainResult}

%%%%%%%%%%%%%%%%%%%%%%%%%%%%%%%%%%%%%%%%%
%%%%%%%%%%%%%%%%%%%%%%%%%%%%%%%%%%%%%%%%%
%%%%%%%%%%%%%%%%%%%%%%%%%%%%%%%%%%%%%%%%%

We set for $i\in\N^*$
\begin{align*}
C_i&=
\begin{pmatrix}
\zeta_{d(i-1)+1,d(i-1)+1}& \zeta_{d(i-1)+1,d(i-1)+2} & \cdots &\zeta_{d(i-1)+1,di}\\
            0          & \zeta_{d(i-1)+2,d(i-1)+2} & \cdots &\zeta_{d(i-1)+2,di}\\
         \vdots        &         \ddots          & \ddots &           \vdots        \\
            0          &         \cdots          &    0   & \zeta_{di,di} \\
\end{pmatrix}\\
\intertext{and}
D_i&=
\begin{pmatrix}
{\zeta_{d(i-2)+1,d(i-1)+1}}^\dagger& {\zeta_{d(i-2)+2,d(i-1)+1}}^\dagger & \cdots &{\zeta_{d(i-1),d(i-1)+1}}^\dagger \\
          0        & {\zeta_{d(i-2)+2,d(i-1)+2}}^\dagger & \cdots & {\zeta_{d(i-1),d(i-1)+2}}^\dagger \\
       \vdots      &       \ddots        & \ddots &          \vdots       \\
          0        &       \cdots        &    0   &{\zeta_{d(i-1),di}}^\dagger\\
\end{pmatrix}.
\end{align*}
For all $i\in\N^*$, $C_i$ are $d\times dK$ matrices and $D_i$ are $dK\times d$ matrices. We fix $\zeta_{i,j}$ with $i\leq0$ or $j\leq0$ so that $C_1D_1=\Id_{d,d}$.

We thereby get the following block description of $H$
\[H_{dn}=
\left(\begin{array}{c|c|c|c|c}
   C_1  & D_2^\dagger &      0_{d,dK}      &  \cdots &      0_{d,dK}      \\\hline
    0_{d,dK}   &     C_2     & D_3^\dagger &  \ddots &   \vdots    \\\hline
 \vdots &   \ddots    &   \ddots    &  \ddots &      0_{d,dK}      \\\hline
    0_{d,dK}   &   \cdots    &      0_{d,dK}      & C_n & D_{n+1}^\dagger \\
\end{array}\right),\]
where $0_{d,dK}$ is the $d\times dK$ zero matrix.

Under the hypothesis (H\ref{moments}), in order to study the limit in $m$ of $Cap_m(\rho)$, it is enough to study $Cap_{nd}(\rho)$ (see Remark \ref{perturbation} following the proof of Lemma \ref{thouless-strip}). We get the following block representation of $G_{dn}$:
\[G_{dn}=\left(\begin{array}{c|c|c|c}
  C_1C_1^\dagger+D_2^\dagger D_2+\lambda\Id_d  &  \left(C_2D_2\right)^\dagger   & 0_d  &  0_d     \\\hline
   C_2D_2  &  \ddots  &  \ddots  &  0_d  \\\hline
   0_d   &  \ddots   &   \ddots  & \left(C_{n}D_{n}\right)^\dagger   \\\hline
  0_d      &   0_d   &  C_{n}D_{n} & C_{n}C_{n}^\dagger+D_{n+1}^\dagger D_{n+1}+\lambda\Id_d \\
\end{array}\right).\]

Note that under (H\ref{frontier}), for all $i\in\N^*$, $C_i D_i$ is a $d\times d$ invertible matrix.

For $i\in\N^*$, we denote by $M_i$ the following matrix
\[\left(\begin{array}{c|c}
0_d&\Id_d\\\hline
-(C_{i+1}D_{i+1})^{-1\dagger}C_{i} D_i&-(C_{i+1} D_{i+1})^{-1\dagger}\left(C_i C_i^\dagger+D_{i+1}^\dagger D_{i+1}+\lambda\Id_d\right)
\end{array}\right)\]
and denote $N_i=\bigwedge^d M_i$. Moreover, $\gamma(N)$ denotes the top Lyapunov exponent associated with $\{N_i\}$, i.e.
\[\gamma(N)\triangleq\lim_{n\rightarrow\infty}\frac{1}{n}\log\nrm{N_n\cdots N_1}.\]
Note that by Theorem \ref{FK}, $\gamma(N)$ is deterministic. See Appendix \ref{sec:LyapunovExponentsTheory} for the definitions concerning the Lyapunov exponents and Appendix \ref{sec:ExteriorProduct} for the relevant background on exterior products. Recall that $(M_i)_{i\in\N^*}$ and $(N_i)_{i\in\N^*}$ depend on $\lambda$.

\begin{thm}
\label{principalK}
Assume (H\ref{ergodicity}), (H\ref{moments}) and (H\ref{frontier}), and set $\lambda=1/\rho$.
\begin{enumerate}
	\item We have
\[Cap_m(\rho)\xrightarrow[m\rightarrow\infty]{}\log{\rho}+\E_{\pi}\log\abs{\zeta_0\zeta_d^\dagger}+\frac{1}{d}\gamma\left(N\right)\triangleq Cap(\rho),\]
where the expectation is taken such that $(\zeta_0,\dots,\zeta_d)$ is distributed according to $\pi$.
	\item As $\rho$ goes to infinity,
\[Cap(\rho)=\log{\rho}+\E_{\pi}\log\abs{\zeta_0\zeta_d^\dagger}+\frac{1}{d}\gamma\left(N(\lambda=0)\right)+o(1).\]
\end{enumerate}
\end{thm}

The theorem is proved in Appendix \ref{sec:RandomSchrOdingerOperatorsTechniques}.

As an alternative to deriving exact analytical results we will also be interested in extracting parameters that characterize the channel rate in the high-SNR regime \cite{Lozano-Tulino-Verdu-high-SNR-IT05}; such parameters are the high-SNR slope (also referred to as the ``multiplexing gain") 
\[\Sinf\triangleq\lim_{\rho\rightarrow\infty} \frac{Cap(\rho)}{\log (K\rho)},\]
and the high-SNR power offset 
\[\Linf\triangleq\lim_{\rho\rightarrow \infty} \frac{1}{\log 2}\left(\log (K\rho)-\frac{Cap(\rho)}{\Sinf}\right),\]
yielding the following affine capacity approximation
\[Cap(\rho) \approx \frac{\Sinf\log2}{3|_\text{dB}}\left(K\rho|_\text{dB} - 3|_\text{dB}\Linf\right).\]

A direct consequence of Theorem \ref{principalK} is the following high-SNR characterization.

\begin{cor}
\label{high-SNR}
Assume (H\ref{ergodicity}), (H\ref{moments}) and (H\ref{frontier}). Then $\Sinf=1$ and
\[\Linf=\frac{1}{\log2}\left[\log K-\E_{\pi}\log\abs{\zeta_0\zeta_d^\dagger}-\frac{1}{d}\gamma(N(\lambda=0))\right].\]
\end{cor}

%%%%%%%%%%%%%%%%%%%%%%%%%%%%%%%%%%%%%%%%%
%%%%%%%%%%%%%%%%%%%%%%%%%%%%%%%%%%%%%%%%%
%%%%%%%%%%%%%%%%%%%%%%%%%%%%%%%%%%%%%%%%%

\section{Reformulations}
\label{sec:Reformulations}

%%%%%%%%%%%%%%%%%%%%%%%%%%%%%%%%%%%%%%%%%
%%%%%%%%%%%%%%%%%%%%%%%%%%%%%%%%%%%%%%%%%
%%%%%%%%%%%%%%%%%%%%%%%%%%%%%%%%%%%%%%%%%

We now derive alternative formulations for $\gamma(N)$ in Subsection \ref{sec:NonAsymptoticResults} and for $\gamma(N(\lambda=0))$ (which characterizes the hign-SNR regime), in Subsection \ref{sec:ResultsInHighSNRRegime}.

%%%%%%%%%%%%%%%%%%%%%%%%%%%%%%%%%%%%%%%%%
%%%%%%%%%%%%%%%%%%%%%%%%%%%%%%%%%%%%%%%%%
%%%%%%%%%%%%%%%%%%%%%%%%%%%%%%%%%%%%%%%%%

\subsection{Non-asymptotic results}
\label{sec:NonAsymptoticResults}

%%%%%%%%%%%%%%%%%%%%%%%%%%%%%%%%%%%%%%%%%
%%%%%%%%%%%%%%%%%%%%%%%%%%%%%%%%%%%%%%%%%
%%%%%%%%%%%%%%%%%%%%%%%%%%%%%%%%%%%%%%%%%

In order to study $\gamma(N)$, we express it as the Lyapunov exponent of simpler matrices. For $i\geq d+1$, we define the following random matrices.
\[\mathfrak{m}_i=\begin{pmatrix}
0&1&0&0\\
\vdots&\ddots&\ddots&0\\
0&\cdots&0&1\\
-\frac{\widetilde{\zeta}_{i,i-d}}{\widetilde{\zeta}_{i,i+d}}&\cdots&\cdots&-\frac{\widetilde{\zeta}_{i,i+d-1}}{\widetilde{\zeta}_{i,i+d}}
\end{pmatrix},\]
where $\widetilde{\zeta}_{i,l}$ is the coefficient in position $(i,l)$ in $G_{dn}$, and set
\[\mathfrak{n}_i=\bigwedge^d\mathfrak{m}_i.\]
Note that $(\mathfrak{m}_i)_{i\geq d+1}$ and $(\mathfrak{n}_i)_{i\geq d+1}$ depend on $\lambda$. We get the following proposition, whose proof is given in Appendix \ref{sec:ProofOfPropositionRefThoulessLike}.

\begin{prop}
\label{thouless-like}
Assume (H\ref{ergodicity}), (H\ref{moments}) and (H\ref{frontier}). Then, $N_i=\mathfrak{n}_{id}\cdots\mathfrak{n}_{(i-1)d+1}$. Therefore, for every $\lambda\geq 0$, $\gamma(N)=d\gamma(\mathfrak{n})$, hence,
	\[Cap(\rho)=\log\rho+\E_{\pi_0,\pi_d}\log\abs{\zeta_0\zeta_d^\dagger}+\gamma\left(\mathfrak{n}\right).\]
\end{prop}

Note that for a given $i\in\N$, $N_i$ depends on $\zeta^{d(i-1)+1},\dots,\zeta^{(d+1)i}$, that is, the fading coefficients of $2d$ different cells. We now want to reduce the product of the $N_i$ to a product of random matrices (that we denote by $\Xi_i$) depending on the fading coefficients of only $d$ cells. Then we reduce it further to a product of random matrices (that we denote by $\xi_i$) depending on the fading coefficients of only one cell.

By doing so, we achieve two goals: first, we express $\gamma(N)$ as the Lyapunov exponent of simpler matrices. Second, if the fading coefficients are i.i.d for different cells, then the $\Xi_i$ and the $\xi_i$ are i.i.d. Products of i.i.d random matrices have been studied extensively (see for example \cite{carmona-lacroix}), moreover, their study can be reduced to the study of a Markov chain on an appropriate space, which can lead to actual analytic expressions (see \cite{bidiagonal} for an example of study of such a Markov chain).

For $i\in\N^*$, we denote by $\Delta_i$ the following matrix
	\[\left(\begin{array}{c|c}
-\left(C_iC_i^{\dagger}+\lambda\Id_d\right)(C_iD_i)^{-1\dagger}&
-C_iD_i+\left(C_iC_i^\dagger+\lambda\Id_d\right) \left(C_iD_i\right)^{-1\dagger}D_i^\dagger D_i\\\hline
(C_iD_i)^{-1\dagger}&-\left(C_iD_i\right)^{-1\dagger}D_i^\dagger D_i
\end{array}\right)\]
and define $\Xi_i=\bigwedge^d\Delta_i$.

For $i\geq d+1$, we denote by $\delta_i$ the following matrix
\[\left(
\begin{array}{c|ccccccc|c}
	\begin{array}{c}
		-\frac{\zeta_{i-d+1,i}}{\zeta_{i-d,i}}\\
		\vdots\\
		-\frac{\zeta_{i-1,i}}{\zeta_{i-d,i}}
	\end{array}& & & &
	\Id_{d-1}& & & &
	0_{d-1,d}\\\hline
	-\frac{\lambda+\abs{\zeta_{i,i}}^2}{\zeta_{i-d,i}\zeta_{i,i}^\dagger}& & & &
	0_{1,d-1}& & & &
	\begin{array}{c c c}
		\lambda\frac{\zeta_{i-d,i}^\dagger}{\zeta_{i,i}^\dagger}&\cdots&\lambda\frac{\zeta_{i-1,i}^\dagger}{\zeta_{i,i}^\dagger}
	\end{array}\\\hline
	\begin{array}{c} \\ \\0_{d-1,1} \\ \\ \\ \end{array}& & & &
	0_{d-1,d-1}& & & &
	\setlength{\arraycolsep}{0pt}
	\begin{array}{c|@{\hspace{1cm}}c@{\hspace{1cm}}} \\ \\
		0_{d-1,1}&\Id_{d-1}\\ \\ \\  
	\end{array}\\\hline
	\frac{1}{\zeta_{i-d,i}\zeta_{i,i}^\dagger}& & & &
	0_{1,d-1}& & & &
	\begin{array}{c c c}
		-\frac{\zeta_{i-d,i}^\dagger}{\zeta_{i,i}^\dagger}&\cdots&-\frac{\zeta_{i-1,i}^\dagger}{\zeta_{i,i}^\dagger}
	\end{array}
\end{array}\right)\]
and define $\xi_i=\bigwedge^d\delta_i$. Note that $(\Delta_i)_{i\in\N^*}$, $(\Xi_i)_{i\in\N^*}$, $(\delta_i)_{i\in\N^*}$ and $(\xi_i)_{i\in\N^*}$ depend on $\lambda$.

\begin{prop}
\label{principalKindep}
Assume (H\ref{ergodicity}), (H\ref{moments}) and (H\ref{frontier}).
\begin{enumerate}
	\item For every $\lambda\geq 0$, $\gamma(\Xi)=\gamma(N)$, hence,
	\[Cap(\rho)=\log\rho+\E_{\pi_0,\pi_d}\log\abs{\zeta_0\zeta_d^\dagger}+\frac{1}{d}\gamma\left(\Xi\right).\]
	\item Assume $K=1$. Then,
	\[\Delta_i=\delta_{id}\cdots\delta_{(i-1)d+1}.\]
	Therefore, for every $\lambda\geq 0$, $\gamma(N)=d\gamma(\xi)$, hence,
	\[Cap(\rho)=\log\rho+\E_{\pi_0,\pi_d}\log\abs{\zeta_0\zeta_d^\dagger}+\gamma\left(\xi\right).\]
\end{enumerate}
\end{prop}

\begin{rmk}
Note that for $K=1$, for all $i\in\N^*$,
\[\left(C_iD_i\right)^{-1\dagger}=C_i^{-1\dagger}D_i^{-1\dagger},\]
and therefore,
\[\Delta_i=\left(\begin{array}{c|c}
-C_iD_i^{-1\dagger}-\lambda(C_iD_i)^{-1\dagger}&\lambda C_i^{-1\dagger}D_i\\\hline
(C_iD_i)^{-1\dagger}&-C_i^{-1\dagger}D_i
\end{array}\right).\]
\end{rmk}

\begin{proof}[Proof of Proposition \ref{principalKindep}]
Let us start by proving point 1. We define for $i\in\N^*$,
\[P_1(i)=\begin{pmatrix}
-C_iD_i&-C_iC_i^\dagger-\lambda\Id_d\\
0_d&\Id_d
\end{pmatrix}\]
and
\[P_2(i)=\begin{pmatrix}
0_d&\Id_d\\
\left(C_iD_i\right)^{-1\dagger}&-\left(C_iD_i\right)^{-1\dagger}D_i^\dagger D_i
\end{pmatrix},\]
so that for all $i\in\N^*$, $M_i=P_2(i+1)P_1(i)$. For $i\in\N^*$, $\Delta_i$ is defined so that $\Delta_i=P_1(i)P_2(i)$. Then, for all $n\in\N^*$,
\begin{align*}
M_n\dots M_1&=P_2(n+1)P_1(n)P_2(n)P_1(n-1)\cdots P_2(2)P_1(1)\\
&=P_2(n+1)\Delta_n\cdots\Delta_2P_1(1).
\end{align*}
and
\begin{align*}
\nrm{N_n\dots N_1}&=\nrm{\bigwedge^d P_2(n+1)\Xi_n\cdots\Xi_2\bigwedge^d P_1(1)}\\
&\leq\nrm{\bigwedge^d P_2(n+1)}\nrm{\Xi_n\cdots\Xi_2}\nrm{\bigwedge^d P_1(1)}.
\end{align*}
Therefore, $\gamma(N)\leq\gamma(\Xi)$. Since $P_1(1)$ and $P_2(n+1)$ are invertible, we get the opposite inequality and point 1 is proved.

The proof of point 2 is postponed to Appendix \ref{sec:ProofOfPropositionRefPrincipalKindep2}.
\end{proof}

%%%%%%%%%%%%%%%%%%%%%%%%%%%%%%%%%%%%%%%%%
%%%%%%%%%%%%%%%%%%%%%%%%%%%%%%%%%%%%%%%%%
%%%%%%%%%%%%%%%%%%%%%%%%%%%%%%%%%%%%%%%%%

\subsection{Results in high-SNR regime}
\label{sec:ResultsInHighSNRRegime}

%%%%%%%%%%%%%%%%%%%%%%%%%%%%%%%%%%%%%%%%%
%%%%%%%%%%%%%%%%%%%%%%%%%%%%%%%%%%%%%%%%%
%%%%%%%%%%%%%%%%%%%%%%%%%%%%%%%%%%%%%%%%%
\begin{prop}
\label{highSNR}
Assume (H\ref{ergodicity}), (H\ref{moments}) and (H\ref{frontier}). Assume moreover that $K=1$.
\begin{enumerate}
	\item For $i\in\N^*$, we set $\Psi_i^1=C_iD_i^{-1\dagger}$ and $\Psi_i^2=C_i^{-1\dagger}D_i $. Then
	\[\Linf=\frac{-1}{\log2}\left[\E_{\pi}\log\abs{\zeta_0\zeta_d^\dagger}+\frac{1}{d}\max_{0\leq i\leq d}\left(\gamma\left(\bigwedge^i \Psi^1\right)+\gamma\left(\bigwedge^{d-i} \Psi^2\right)\right)\right].\]
		\item For $i\geq d+1$, we set\setlength{\arraycolsep}{0pt}
\[\psi_i^1=\left(
\begin{array}{@{\hspace{2pt}}c@{\hspace{2pt}}|c}
	\begin{array}{c}
		-\frac{\zeta_{i-d+1,i}}{\zeta_{i-d,i}}\\
		\vdots\\
		-\frac{\zeta_{i,i}}{\zeta_{i-d,i}}		
	\end{array}&\setlength{\arraycolsep}{12pt}
	\begin{array}{c}
	\\
		\Id_{d-1}\\
				\\\hline

		0_{1,d-1}
	\end{array}
\end{array}\right)\textrm{ and }
\psi_i^2=\left(
\begin{array}{@{\hspace{2pt}}c@{\hspace{2pt}}|c}
	\begin{array}{c}
		-\frac{\zeta_{i-1,i}}{\zeta_{i,i}}\\
		\vdots\\
		-\frac{\zeta_{i-d,i}}{\zeta_{i,i}}		
	\end{array}&\setlength{\arraycolsep}{12pt}
	\begin{array}{c}
	\\
		\Id_{d-1}\\
				\\\hline

		0_{1,d-1}
	\end{array}
\end{array}\right)^\dagger
.\]
\setlength{\arraycolsep}{5pt}
Then,
	\[\Linf=\frac{-1}{\log2}\left[\E_{\pi}\log\abs{\zeta_0\zeta_d^\dagger}+\max_{0\leq i\leq d}\left(\gamma\left(\bigwedge^i \psi^1\right)+\gamma\left(\bigwedge^{d-i} \psi^2\right)\right)\right].\]
\end{enumerate}
\end{prop}

\begin{rmk}
\begin{enumerate}
	\item Recall that for a stationary ergodic sequence of complex random matrices $(X_i)_{i\in\N^*}$ of size $d$,
	\[\gamma\left(\bigwedge^0 X\right)=0\textrm{ and }\gamma\left(\bigwedge^d X\right)=\E\log\abs{\det X_1}.\] 
	\item Note that if for $0\leq i\leq d$, $\pi_i=\pi_{d-i}$ and the vectors $(\zeta_i)_{i\in\N^*}$ are i.i.d, then $(\psi_i^1)_{i\geq d+1}$ and $((\psi_i^2)^\dagger)_{i\geq d+1}$ have the same distribution and $(\psi_i^2)_{i\geq d+1}$ and $((\psi_i^2)^\dagger)_{i\geq d+1}$ have the same Lyapunov exponents. Therefore, as $\rho$ goes to infinity,
\[\Linf=\frac{-1}{\log2}\left[\E_{\pi}\log\abs{\zeta_0\zeta_d^\dagger}+\gamma\left(\bigwedge^{\left\lceil d/2\right\rceil} \psi^1\right)+\gamma\left(\bigwedge^{\left\lfloor d/2\right\rfloor} \psi^1\right)\right].\]
\end{enumerate}
\end{rmk}

\begin{proof}[Proof of Proposition \ref{highSNR}]
Using Corollary \ref{high-SNR} and Proposition \ref{principalKindep}, in order to prove point 1, we only have to prove that
\begin{equation}
\label{goal1}
\gamma(\Xi(\lambda=0))=\max_{0\leq i\leq d}\left(\gamma\left(\bigwedge^i \Psi^1\right)+\gamma\left(\bigwedge^{d-i} \Psi^2\right)\right).
\end{equation}
Recall that $\gamma(\Xi)=\gamma_1(\Delta)+\cdots+\gamma_d(\Delta)$ and that 	
\[\Delta_i(\lambda=0)=\left(\begin{array}{c|c}
-\Psi_i^1&0_{d,d}\\\hline
(C_iD_i)^{-1\dagger}&-\Psi_i^2
\end{array}\right).\]

By Proposition \ref{reductible}, the sequence $\gamma_1(\Delta(\lambda=0)),\dots,\gamma_{2d}(\Delta(\lambda=0))$ is equal up to the order to the sequence \[\gamma_1(\Psi^1),\dots,\gamma_d(\Psi^1),\gamma_1(\Psi^2),\dots,\gamma_{d}(\Psi^2).\]
Therefore, (\ref{goal1}) is a direct consequence of $\gamma(\Xi(\lambda=0))=\gamma_1(\Delta(\lambda=0))+\cdots+\gamma_{d}(\Delta(\lambda=0))$ and $\gamma\left(\bigwedge^i \Psi^{1,2}\right)=\gamma_1(\Psi^{1,2})+\cdots+\gamma_i(\Psi^{1,2})$.

The proof of point 2 goes along the same lines using the fact that
\[\delta_i(\lambda=0)=\left(\begin{array}{c|c}
-\psi_i^1&0_{d,d}\\\hline
\psi_i^3&-\widetilde{\psi}_i^2
\end{array}\right),\]
where
\[\widetilde{\psi}_i^2=\begin{pmatrix}0&0&1\\0&\iddots&0\\1&0&0\end{pmatrix}\psi_i^2\begin{pmatrix}0&0&1\\0&\iddots&0\\1&0&0\end{pmatrix}\textrm{ and }
\tilde{\psi}_i^3=\left(\begin{array}{c|c}0_{d-1,1}&0_{d-1,d-1}\\\hline\frac{1}{\zeta_{i-d,i}\zeta_{i,i}^\dagger}&0_{1,d-1}\end{array}\right),\]
therefore, the Lyapunov exponents of $\widetilde{\psi}^2$ and $\psi^2$ are the same.
\end{proof}

%%%%%%%%%%%%%%%%%%%%%%%%%%%%%%%%%%%%%%%%%
%%%%%%%%%%%%%%%%%%%%%%%%%%%%%%%%%%%%%%%%%
%%%%%%%%%%%%%%%%%%%%%%%%%%%%%%%%%%%%%%%%%

\section{Bounds on the capacity}
\label{sec:BoundsOnTheCapacity}

%%%%%%%%%%%%%%%%%%%%%%%%%%%%%%%%%%%%%%%%%
%%%%%%%%%%%%%%%%%%%%%%%%%%%%%%%%%%%%%%%%%
%%%%%%%%%%%%%%%%%%%%%%%%%%%%%%%%%%%%%%%%%

%%%%%%%%%%%%%%%%%%%%%%%%%%%%%%%%%%%%%%%%%
%%%%%%%%%%%%%%%%%%%%%%%%%%%%%%%%%%%%%%%%%
%%%%%%%%%%%%%%%%%%%%%%%%%%%%%%%%%%%%%%%%%

\subsection{Bounds on the top Lyapunov exponent}
\label{sec:BoundsOnTheTopLyapunovExponent}

%%%%%%%%%%%%%%%%%%%%%%%%%%%%%%%%%%%%%%%%%
%%%%%%%%%%%%%%%%%%%%%%%%%%%%%%%%%%%%%%%%%
%%%%%%%%%%%%%%%%%%%%%%%%%%%%%%%%%%%%%%%%%

We use the Fr\"obenius norm on the matrices, it is a sub-multiplicative norm, therefore, we can apply (\ref{subadditive}) to the different formulations of the capacity to get the following proposition.

\begin{prop}
\label{bound-lyapunov}
Assume (H\ref{ergodicity}), (H\ref{moments}) and (H\ref{frontier}).
\begin{enumerate}
	\item\label{nstepgeneral} For $\lambda=1/\rho$ and $p\geq1$,
\[Cap(\rho)\leq\log\rho+\E_{\pi_0,\pi_d}\log\abs{\zeta_0\zeta_d^\dagger}+
\frac{1}{dp}\E\log\nrm{N_p(\lambda)\cdots N_{1}(\lambda)}.\]
	\item\label{nstepthouless} For $\lambda=1/\rho$ and $p\geq1$,
\[Cap(\rho)\leq\log\rho+\E_{\pi_0,\pi_d}\log\abs{\zeta_0\zeta_d^\dagger}+
\frac{1}{p}\E\log\nrm{\mathfrak{n}_p(\lambda)\cdots \mathfrak{n}_{1}(\lambda)}.\]
	\item\label{nstepindep} For $\lambda=1/\rho$ and $p\geq1$,
\[Cap(\rho)\leq\log\rho+\E_{\pi_0,\pi_d}\log\abs{\zeta_0\zeta_d^\dagger}+
\frac{1}{dp}\E\log\nrm{\Xi_p(\lambda)\cdots \Xi_{1}(\lambda)}.\]
	\item\label{nstepindepthouless} Assume $K=1$. For $\lambda=1/\rho$ and $p\geq1$,
\[Cap(\rho)\leq\log\rho+\E_{\pi_0,\pi_d}\log\abs{\zeta_0\zeta_d^\dagger}+
\frac{1}{p}\E\log\nrm{\xi_p(\lambda)\cdots \xi_{1}(\lambda)}.\]
\end{enumerate}
Moreover, the bounds are tight as $p$ goes to infinity.
\end{prop}

The bound of point 2 with $p=1$ can be reformulated as follows.

\begin{cor}
\label{1stepthouless} For $\lambda=1/\rho$,
\begin{align*}
&Cap(\rho)\leq\log\rho+\\
&\frac{1}{2}\E\log\Bigg[\binom{2d-2}{d-1}\left(
\sum_{l=2}^{2d}\abs{\sum_{t=(l-d-1)\vee0}^{(l-1)\wedge d}\zeta_{d+1,d+t+1}\zeta_{l,d+t+1}^\dagger+\lambda\1_{(l=d+1)}}^2
\right)+\\
&\binom{2d-1}{d}\left(\abs{\zeta_{d+1,d+1}\zeta_{1,d+1}^\dagger}^2+\abs{\zeta_{d+1,2d+1}\zeta_{2d+1,2d+1}^\dagger}^2\right)\Bigg].
\end{align*}
\end{cor}
The proof is postponed to Appendix \ref{sec:ProofOfCorollaryRef1stepthouless}.

%%%%%%%%%%%%%%%%%%%%%%%%%%%%%%%%%%%%%%%%%
%%%%%%%%%%%%%%%%%%%%%%%%%%%%%%%%%%%%%%%%%
%%%%%%%%%%%%%%%%%%%%%%%%%%%%%%%%%%%%%%%%%

\subsection{Other bounds}
\label{sec:OtherBounds}

%%%%%%%%%%%%%%%%%%%%%%%%%%%%%%%%%%%%%%%%%
%%%%%%%%%%%%%%%%%%%%%%%%%%%%%%%%%%%%%%%%%
%%%%%%%%%%%%%%%%%%%%%%%%%%%%%%%%%%%%%%%%%

\begin{prop}
\label{information}
Assume (H\ref{ergodicity}) and (H\ref{moments}). For $\lambda=1/\rho$,
\begin{align*}	&\max\left(\E_{\pi_0}\log\left(\lambda+\abs{\zeta_0}^2\right),\E_{\pi_d}\log\left(\lambda+\abs{\zeta_d}^2\right)\right)\leq Cap(\rho)-\log\rho\leq\\
	&\E\log\left(\lambda+\abs{\zeta_{1,1}}^2+\cdots+\abs{\zeta_{1,d+1}}^2\right).
\end{align*}
\end{prop}

\begin{proof}
The upper bound is a consequence of Hadamard's inequality for semi-positive definite Hermitian matrices. Indeed,
\[\frac{1}{m}\log\det G_m(\lambda)\leq\frac{1}{m}\sum_{i=1}^m\log\left(\lambda+\abs{\zeta_{i,i}}^2+\cdots+\abs{\zeta_{i,i+d}}^2\right).\]
Let us show the lower bound of point \ref{information} using the tools of \cite{SOW}.
\begin{align*}
Cap_m(\rho)&=\frac{1}{m}I\left(\vect{x},\vect{y}|(\zeta_{i,i})_{1\leq i\leq m},\dots,(\zeta_{i,i+d})_{1\leq i\leq m}\right)\\
&=\frac{1}{m}\sum_{j=1}^m I\left(x_j,\vect{y}|(x_i)_{1\leq i<j},(\zeta_{i,i})_{1\leq i\leq m},\dots,(\zeta_{i,i+d})_{1\leq i\leq m}\right)\\
&\geq\frac{1}{m}\sum_{j=1}^m I\left(x_j,y_{j-d}|(x_i)_{1\leq i<j},(\zeta_{i,i})_{1\leq i\leq m},\dots,(\zeta_{i,i+d})_{1\leq i\leq m}\right)\\
&=\frac{1}{m}\sum_{j=1}^m I\left(x_j,\zeta_{j-d,j}x_j+n_{j-d}|\zeta_{j-d,j}\right),
\end{align*}
which is the per-cell sum-rate capacity of a single user fading channel. Therefore, the lower bound is $\E_{\pi_d}\log\left(1+\rho\abs{\zeta_d}^2\right)$.

The role of the distributions $\pi_0$ and $\pi_d$ can be exchanged by a right-left reflection, namely the transformation $\zeta_{i,j}'=\zeta_{m-i+1,m+d-j+1}$. Thus, we get the lower bound.
\end{proof}

In the end of this section, we slightly modify the setting, by considering $m$ cells with $K$ single antenna users per cell and $m+d$ single antenna BSts. The communication is characterized by the following $(m+d)\times Km$ channel transfer matrix $H_m$, which is a $d+1$ block diagonal matrix defined by,
\[H_m=
\begin{pmatrix}
\zeta_{1,1}& 0 & \cdots & 0    \\
   \zeta_{2,1}     & \zeta_{2,2} &  &\vdots \\
 \vdots  &    \vdots       & \ddots &  0 \\
     \zeta_{d+1,1}   &   \zeta_{d+1,2}  &       & \zeta_{m,m} \\  
  0  & \zeta_{d+2,2} &   & \zeta_{m+1,m}\\
 \vdots &    & \ddots & \vdots\\
 0 &  \cdots  &  0 &  	\zeta_{d+m,m}
\end{pmatrix},\]
where $\zeta_{i,j}$ are $1\times K$ row vectors.

We consider the per-cell sum-rate capacity $Cap_m(\rho)$ that is given by (\ref{capacity}). We denote by $Cap(\rho)$ the limit of $Cap_{m}(\rho)$ as $m$ tends to infinity.

Note that in the limit, this setting is equivalent to the setting we define in Section \ref{sec:ProblemStatement}. In particular, the normalization by $1/m$ or $1/(m+d)$ is equivalent.

\begin{prop}
\label{truncation}
For all $n\in\N^*$,
\[\frac{n}{n+d}Cap_n\left(\frac{n+d}{n}\rho\right)\leq Cap(\rho)\leq Cap_n(\rho).\]
Moreover, the bounds are tight as $n$ goes to infinity.
\end{prop}

Note that taking the upper bound for $n=1$, one gets the upper bound of Proposition \ref{information}.

\begin{figure}
\begin{center}
\includegraphics[scale=0.55]{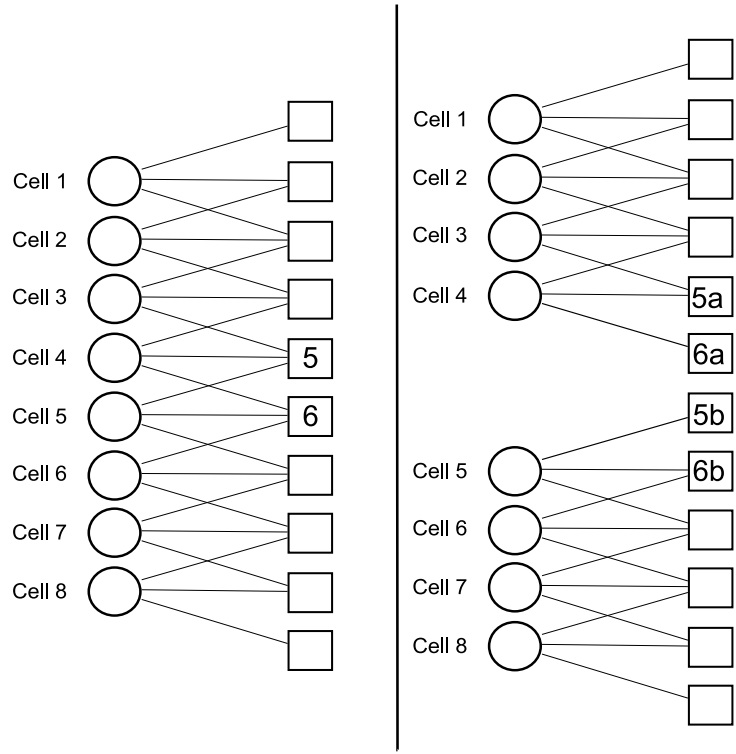}
\end{center}
\caption{Proof of Proposition \ref{truncation}, upper bound}
\label{fig: upper_bound}
\end{figure}

\begin{figure}
\begin{center}
\includegraphics[scale=0.55]{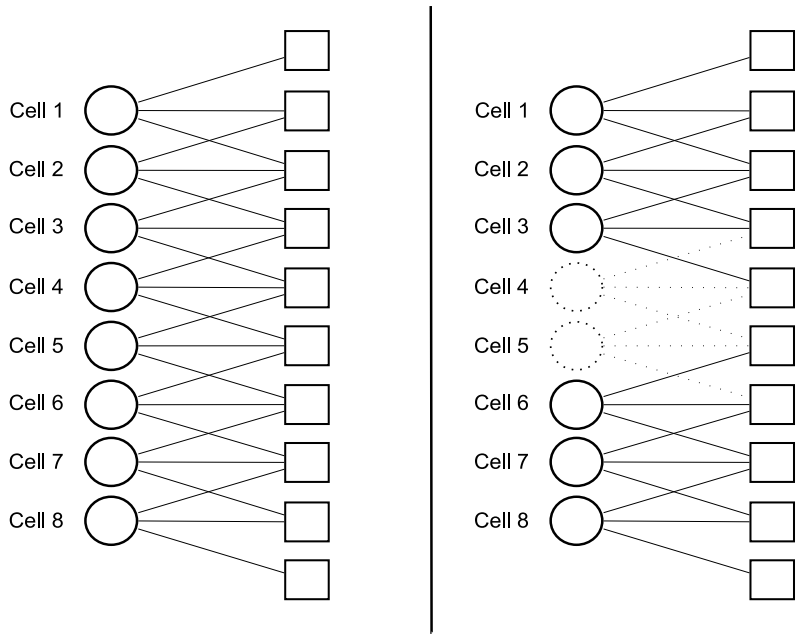}
\end{center}
\caption{Proof of Proposition \ref{truncation}, lower bound}
\label{fig: lower_bound}
\end{figure}

\begin{proof}
Number the cells from 1 to $m$ and the antennas from 1 to  $m+d$.

\emph{Upper bound}. Take $1\leq n\leq m$. Consider the following transformation of the communication system: duplicate the antennas number $n+1$ to $n+d$ (each antenna is replaced by two antennas) and denote by $(n+1,a), (n+1,b), \dots, (n+d,a), (n+d,b)$ the new antennas. The cells number $n-d+1$ to $n$ broadcast toward the antennas $(n+1,a), (n+2,a), \dots, (n+d,a)$ and the cells number $n+1$ to $n+d$ broadcast toward the antennas $(n+1,b), (n+2,b), \dots, (n+d,b)$. See Figure \ref{fig: upper_bound} for an illustration with $d=2$, $m=8$ and $n=4$. The new system has a higher capacity since a first step of the decoding could be summing up the signals received at antennas $(n+i,a)$ and $(n+i,b)$ ($1\leq i\leq d$) to get the signal received at antenna $n+i$ in the former system. Therefore,
\[m Cap_m(\rho)\leq n Cap_n(\rho) + (m-n) Cap_{m-n}(\rho).\]
Take $k\in\N^*$, by induction,
\[(nk) Cap_{nk}(\rho)\leq (nk) Cap_{n}(\rho).\]
Dividing by $nk$ and taking $k$ to infinity in the LHS gives the upper bound.

\emph{Lower bound}. Take $k\in\N^*$ and consider $k(n+d)$ users and their corresponding antennas. For every group of $n+d$ users, silence the $d$ last users and redistribute their power to the $n$ first users so that their SNR becomes $(n+d/n) \rho$ (the average SNR is still $\rho$). See Figure \ref{fig: lower_bound} for an illustration. Since asymptotically, the equal power distribution among the users is optimal \cite[Appendix C]{Tulino-Lozano-Verdu}, we get that the new system has a lower capacity. Therefore, for $k$ going to infinity,
\[(nk) Cap_n\left(\frac{n+d}{n} \rho\right)\leq k(n+d)Cap_{k(n+d)}(\rho).\]
Dividing by $k(n+d)$ and taking $k$ to infinity in the LHS gives the lower bound.
\end{proof}

%%%%%%%%%%%%%%%%%%%%%%%%%%%%%%%%%%%%%%%%%
%%%%%%%%%%%%%%%%%%%%%%%%%%%%%%%%%%%%%%%%%
%%%%%%%%%%%%%%%%%%%%%%%%%%%%%%%%%%%%%%%%%

\subsection{Numerical comparison of the bounds}
\label{sec:NumericalComparaisonOfTheBounds}

%%%%%%%%%%%%%%%%%%%%%%%%%%%%%%%%%%%%%%%%%
%%%%%%%%%%%%%%%%%%%%%%%%%%%%%%%%%%%%%%%%%
%%%%%%%%%%%%%%%%%%%%%%%%%%%%%%%%%%%%%%%%%

We first compare the bound of Corollary \ref{1stepthouless} and the bounds of Proposition \ref{information}. Note that by the ergodic theorem, the upper-bound of Proposition \ref{information} grows like $\log d$, whereas in the bound of Corollary \ref{1stepthouless}, the part $\log\binom{2d-1}{d}\abs{\zeta_{d+1,d+1}\zeta_{1,d+1}^\dagger}^2$ alone already grows like $d$. Nevertheless, it is not necessarily true that the upper-bound of Proposition \ref{information} is better that the one of Corollary \ref{1stepthouless} for all $d$ and all fading distributions.

\begin{figure}
\begin{center}
\psfrag{TTTTheoreminformationLB}{\scriptsize Proposition \ref{information} (LB)}
\psfrag{TheoreminformationUB}{\scriptsize Proposition \ref{information} (UB)}
\psfrag{Theoremgeneral}{\scriptsize Corollary \ref{1stepthouless}}
\includegraphics[scale=0.71]{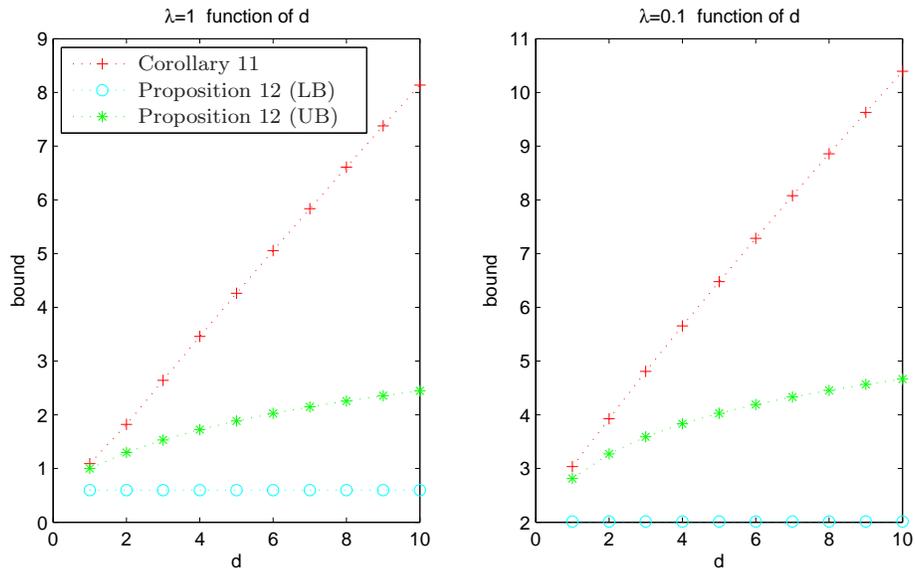}
\end{center}
\caption{Comparison of the bounds of Corollary \ref{1stepthouless} and Proposition \ref{information} for $\lambda=0.1,1$, in function of $d$.}
\label{fig: 1step}
\end{figure}

In Figure \ref{fig: 1step}, we present the bounds of Corollary \ref{1stepthouless} and Proposition \ref{information} in the special case of Rayleigh fading (real and imaginary parts are independent Gaussian random variables with zero mean and variance $1/\sqrt{2}$). The curves are produced by Monte Carlo simulation with $10^6$ samples. We see that in this case, even for $d$ small, the upper-bound of Proposition \ref{information} is better than the one of Corollary \ref{1stepthouless}.

\begin{figure}
\begin{center}
\psfrag{TheoreminformationLBBBBB}{\scriptsize Proposition \ref{information} (LB)}
\psfrag{TheoreminformationUB}{\scriptsize Proposition \ref{information} (UB)}
\psfrag{Theoremgennsteps}{\scriptsize Proposition \ref{bound-lyapunov}.\ref{nstepgeneral}}
\psfrag{Propositionindnsteps}{\scriptsize Proposition \ref{bound-lyapunov}.\ref{nstepindep}}
\psfrag{coupelower}{\scriptsize Proposition \ref{truncation} (LB)}
\psfrag{coupeupper}{\scriptsize Proposition \ref{truncation} (UB)}
\includegraphics[scale=0.7]{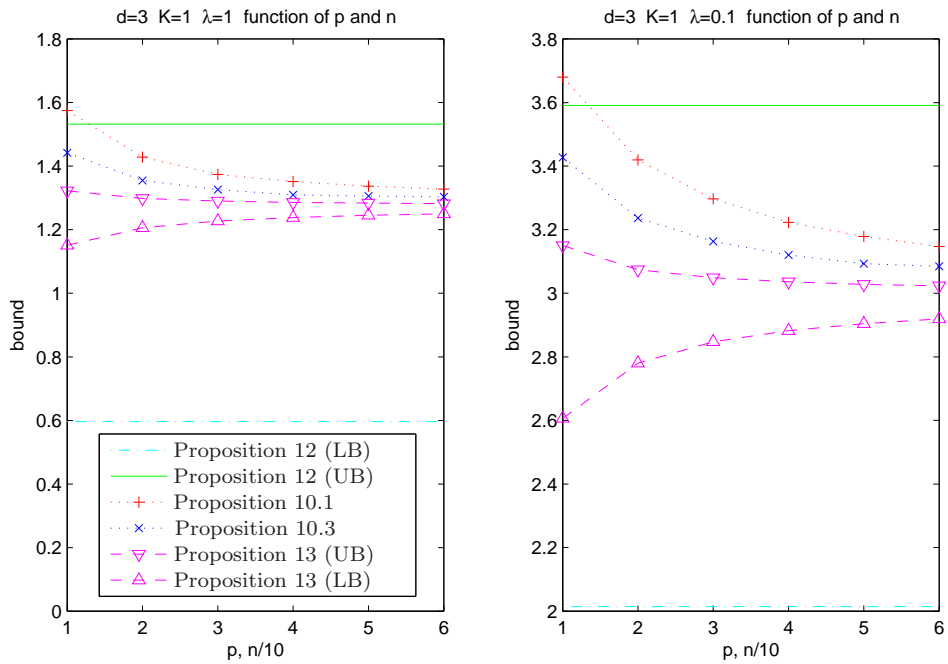}
\end{center}
\vskip-0.5cm
\caption{Comparison of the different bounds for $\lambda=0.1,1$ and $d=3$.}
\label{fig: nsteps_d3}
\end{figure}

\begin{figure}
\begin{center}
\psfrag{TheoreminformationLB}{\scriptsize Proposition \ref{information} (LB)}
\psfrag{TheoreminformationUB}{\scriptsize Proposition \ref{information} (UB)}
\psfrag{Theoremgennsteps}{\scriptsize Proposition \ref{bound-lyapunov}.\ref{nstepgeneral}}
\psfrag{PPPPPPropositionKindKnsteps}{\scriptsize Proposition \ref{bound-lyapunov}.\ref{nstepindep}}
\psfrag{coupelower}{\scriptsize Proposition \ref{truncation} (LB)}
\psfrag{coupeupper}{\scriptsize Proposition \ref{truncation} (UB)}
\includegraphics[scale=0.58]{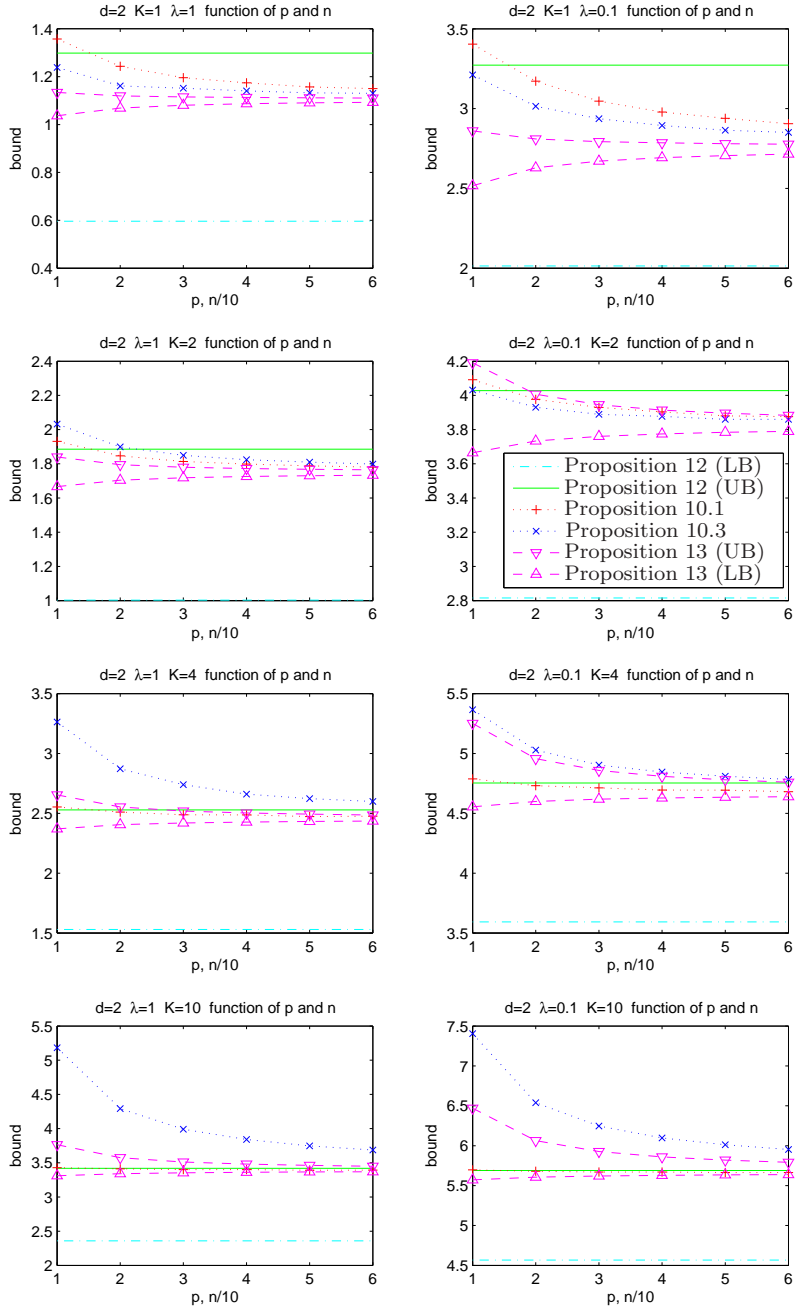}
\end{center}
\vskip-2cm
\caption{Comparison of the different bounds for $\lambda=0.1,1$ and $K=1,2,4,10$. The values for the non random channel with $\lambda=1$ (resp. $\lambda=0.1$) and $K=1,2,4,10$ are respectively $1.06,\ 1.47,\ 1.95,\ 2.66$ (resp. $2,66,\ 3.25,\ 3.87,\ 4.72$).}
\label{fig: nsteps_K_d2}
\end{figure}

In Figures \ref{fig: nsteps_d3} and \ref{fig: nsteps_K_d2}, we compare the bounds of Proposition \ref{bound-lyapunov}, point \ref{nstepgeneral} and \ref{nstepindep}, Proposition \ref{information} and Proposition \ref{truncation} in the special case of Rayleigh fading (real and imaginary parts are independent Gaussian random variables with zero mean and variance $1/\sqrt{2}$). The curves are produced by Monte Carlo simulations with $10^5$ samples.

Note that in the case $d=2$, for $K=1$, the bounds of Proposition \ref{bound-lyapunov}, point \ref{nstepindep} are better than those of point \ref{nstepgeneral}, whereas for $K>1$, it is the opposite.

We see that in the case $d=2$, for $K=4$ and $K=10$, the upper-bound of Proposition \ref{information} is very close to the capacity and the upper-bounds of Proposition \ref{bound-lyapunov}.\ref{nstepgeneral} are getting tight very rapidly.

In the case $d=2$, we want to compare the random-fading channel with the non-fading channel. See Appendix \ref{sec:CapacityOfTheNonFadingChannels} for the per-cell sum-rate capacity of the non-fading channel. The comparison is done in Figure \ref{fig: nsteps_K_d2}; in the eight cases that we consider, the random-fading channel is better than the non random one.

%%%%%%%%%%%%%%%%%%%%%%%%%%%%%%%%%%%%%%%%%
%%%%%%%%%%%%%%%%%%%%%%%%%%%%%%%%%%%%%%%%%
%%%%%%%%%%%%%%%%%%%%%%%%%%%%%%%%%%%%%%%%%

\section{Results for particular cases in the high-SNR regime}
\label{sec:ResultsForParticularCasesInTheHighSNRRegime}

%%%%%%%%%%%%%%%%%%%%%%%%%%%%%%%%%%%%%%%%%
%%%%%%%%%%%%%%%%%%%%%%%%%%%%%%%%%%%%%%%%%
%%%%%%%%%%%%%%%%%%%%%%%%%%%%%%%%%%%%%%%%%

%%%%%%%%%%%%%%%%%%%%%%%%%%%%%%%%%%%%%%%%%
%%%%%%%%%%%%%%%%%%%%%%%%%%%%%%%%%%%%%%%%%
%%%%%%%%%%%%%%%%%%%%%%%%%%%%%%%%%%%%%%%%%

\subsection{Case $d=1$}
\label{sec:CaseD1}

%%%%%%%%%%%%%%%%%%%%%%%%%%%%%%%%%%%%%%%%%
%%%%%%%%%%%%%%%%%%%%%%%%%%%%%%%%%%%%%%%%%
%%%%%%%%%%%%%%%%%%%%%%%%%%%%%%%%%%%%%%%%%

As a direct application of Proposition \ref{highSNR} in the case $d=1$ and $K=1$, we get the following result.

\begin{prop}
Assume (H\ref{ergodicity}), (H\ref{moments}) and (H\ref{frontier}). Then
\[\Linf=\frac{-1}{\log2}\left[2\max\left(\E_{\pi_0}\log\abs{\zeta_0}\,;\,\E_{\pi_1}\log\abs{\zeta_1}\right)\right].\]
\end{prop}

Note that a similar result was already proved by other techniques in \cite{bidiagonal} under much stronger hypothesis, in particular, independence of the fading coefficients was assumed there. In contrary, our result depends only on the marginal distributions of the fading coefficients and is valid for a larger class of joint distributions.

We want to compare the per-cell sum-rate capacity of the random-fading and non-fading channels. For a random variable $\zeta$, by Jensen's inequality,
\[\E\log\abs{\zeta}^2\leq\log\E\abs{\zeta}^2.\]
Therefore, under the constraints $\E_{\pi_0}\abs{\zeta_0}^2\leq1$ and $\E_{\pi_1}\abs{\zeta_1}^2\leq1$, the non-fading channel achieves the best per-cell sum-rate capacity in the high SNR regime.

%%%%%%%%%%%%%%%%%%%%%%%%%%%%%%%%%%%%%%%%%
%%%%%%%%%%%%%%%%%%%%%%%%%%%%%%%%%%%%%%%%%
%%%%%%%%%%%%%%%%%%%%%%%%%%%%%%%%%%%%%%%%%

\subsection{Case $d=2$}
\label{sec:CaseD2}

%%%%%%%%%%%%%%%%%%%%%%%%%%%%%%%%%%%%%%%%%
%%%%%%%%%%%%%%%%%%%%%%%%%%%%%%%%%%%%%%%%%
%%%%%%%%%%%%%%%%%%%%%%%%%%%%%%%%%%%%%%%%%

We now assume that $d=2$ and $K=1$ and that the fading coefficients have the following form; for $i\in\N^*$,
\[\zeta_{i-2,i}=\alpha a_i\ ,\quad\zeta_{i-1,i}=\beta b_i\textrm{ and }\zeta_{i,i}= c_i,\]
where $a_i$, $b_i$ and $c_i$ are random variable distributed according to $\pi_a$, $\pi_b$ and $\pi_c$ respectively and  $\alpha$ and $\beta$ are parameters such that $\alpha>0$ and $\beta\geq0$. Moreover, take the following normalization that can always be achieved by modifying $\alpha$ and $\beta$.
\[\E_{\pi_a}\log\abs{a_1}=\E_{\pi_b}\log\abs{b_1}=\E_{\pi_c}\log\abs{c_1}.\]
We use the notation of Proposition \ref{highSNR}.

\begin{prop}
\label{d2}
Assume that $(a_i,b_i,c_i)_{i\in\N^*}$ is a stationary ergodic sequence such that for all $i\in\N^*$, almost surely, $a_i$ and $c_i$ are non zero and that their exist $\ep>0$ such that $\E_{\pi_a}\left(\log\abs{a_1}\right)^{1+\ep}$, $\E_{\pi_b}\left(\log\abs{b_1}\right)^{1+\ep}$ and $\E_{\pi_c}\left(\log\abs{c_1}\right)^{1+\ep}$ are finite.

Then, there exist a domain ${\cal D}\subset(0,1]\times[0,1]$ such that for all $(x,y)\in{\cal D}$, $(0,x)\times[0,y)\subset{\cal D}$ and for all $(\alpha,\beta)\in{\cal D}$, as $\rho$ goes to infinity,
\begin{equation}
\label{highSNRd2}
\Linf=\frac{-2}{\log2}\E_{\pi_a}\log\abs{a}.
\end{equation}
\end{prop}

The proof is postponed to Appendix \ref{sec:ProofOfPropositionRefD2}.

\begin{rmk}
\begin{enumerate}
	\item The set ${\cal D}$ is not maximal in the sense that (\ref{highSNRd2}) may hold for couples $(\alpha,\beta)\notin{\cal D}$.
	\item Note that for $(\alpha,\beta)\in{\cal D}$, in the high-SNR regime, the lower bound of Proposition \ref{information} is tight.
	\item The proof will yield an \emph{effective} construction of ${\cal D}$, which allows us to find many points in ${\cal D}$. Indeed, we construct $(f_p)_{p\in\N^*}$ a family of functions on $(0,1]\times[0,1]$ with the following property: if there exists $p\in\N^*$ such that $f_p(\alpha,\beta)\leq0$, then (\ref{highSNRd2}) holds.
	\item Note that (\ref{highSNRd2}) does not hold when $\alpha>1$, indeed, as it will appear in the course of the proof, as $\rho$ goes to infinity,
\[Cap(\rho)\geq\log\rho+2\E_{\pi_a}\log\abs{a}+\log\alpha.\]
We conjecture that (\ref{highSNRd2}) does not hold when $\beta>1$ either.
\end{enumerate}
\end{rmk}

\begin{figure}
\begin{center}
\includegraphics[scale=0.55]{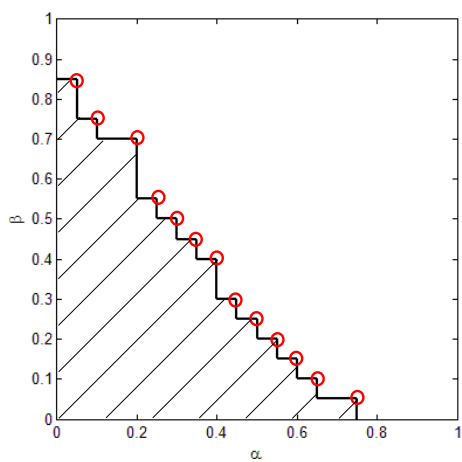}
\end{center}
\caption{Region where (\ref{highSNRd2}) holds for Rayleigh fading.}
\label{fig: D}
\end{figure}

Let us apply Proposition \ref{d2} to the case where $(a_i,b_i,c_i)_{i\in\N^*}$ are independent Rayleigh distributed coefficients. In Figure \ref{fig: D}, we plot points for which $f_{20}$ is less or equal to -0.05 (Monte Carlo simulations realized with $10^5$ samples). Therefore, (\ref{highSNRd2}) holds for $(\alpha,\beta)$ in the stripped region and in particular for $\alpha,\beta\leq0.4$. Note that in this case, the power offset is
\[\Linf=\frac{\gamma}{\log2},\]
where $\gamma$ is the Euler constant.

%%%%%%%%%%%%%%%%%%%%%%%%%%%%%%%%%%%%%%%%%
%%%%%%%%%%%%%%%%%%%%%%%%%%%%%%%%%%%%%%%%%
%%%%%%%%%%%%%%%%%%%%%%%%%%%%%%%%%%%%%%%%%

\subsection{Artificial fading}
\label{sec:ArtificialFading}

%%%%%%%%%%%%%%%%%%%%%%%%%%%%%%%%%%%%%%%%%
%%%%%%%%%%%%%%%%%%%%%%%%%%%%%%%%%%%%%%%%%
%%%%%%%%%%%%%%%%%%%%%%%%%%%%%%%%%%%%%%%%%

In the frame of non-fading channels, we consider artificial fading, that is, every user uses a pseudo-random fading and multiplies its signal by this artificial fading. The fading coefficients then have the following form, for $i\in\N^*$ and $0\leq s\leq d$,
\[\zeta_{i,i+s}=\alpha_s P_{i+s},\]
where $\alpha_0,\dots\alpha_d$ are non random positive numbers and $P_i\,,i\in\N^*$ are stationary ergodic pseudo-random complex row vectors of size $K$ distributed according to a law denoted by $\pi_P$. We moreover assume that for all $i\in\N^*$, almost surely, the coefficients of $P_i$ are non zero and that $\E_{\pi_P}\nrm{P_1}^2=1$.

In \cite{Somekh-Shamai-2000}, it is proved that in the case $d=2$, the per-cell sum-rate capacity is smaller with artificial fading. Indeed, had such a procedure helped, then it would be used in non-fading situations to enhance capacity. It is evident then that it is deleterious, as the expression in Proposition \ref{artificial} exhibits.

We consider the high-SNR regime and derive the explicit influence of the artificial fading.

\begin{prop}
\label{artificial}
Denote by $\Linf^0$ the power off-set without artificial fading (that is, $P_i=(1,\dots,1)$ almost surely) and by $\Linf$ the power off-set with artificial fading. Then,
\[\Linf=\Linf^0-\frac{1}{\log 2}\E_{\pi_P}\log\nrm{P_1}^2.\]
\end{prop}

\begin{rmk}
By Jensen's inequality, we get that $\Linf\geq\Linf^0$, therefore, in the high-SNR regime, the per-cell sum-rate capacity is smaller with artificial fading.
\end{rmk}

\begin{proof}[Proof of Proposition \ref{artificial}]
We set until the end of the proof $\lambda=0$. Using Corollary \ref{high-SNR} and Proposition \ref{principalKindep},
\[\Linf=\frac{1}{\log2}\left[\log K-\E_\pi\log\abs{\alpha_0\alpha_d^\dagger}-\E_{\pi_P}\log\nrm{P_1}^2-\frac{1}{d}\gamma(\Delta)\right],\]
whereas
\[\Linf^0=\frac{1}{\log2}\left[\log K-\E_\pi\log\abs{\alpha_0\alpha_d^\dagger}-\frac{1}{d}\gamma\left(\widetilde{\Delta}\right)\right],\]
where $\left(\widetilde{\Delta}_i\right)_{i\in\N^*}$ denote the matrices without artificial fading. Therefore, we only have to prove that for $i\in\N^*$, $\Delta_i$ does not depend on $(P_i)_{i\in\N^*}$. In the case $K=1$, $\Delta_i=\delta_{id}\cdots\delta_{(i-1)d+1}$ and for $i\geq d+1$, $\delta_i$ does not depend on $(P_i)_{i\in\N^*}$, therefore, $\Delta_i$ does not depend on $(P_i)_{i\in\N^*}$.

Let us assume $K>1$. Using Proposition \ref{principalKindep}, we only have to prove that for $i\in\N^*$, $\Delta_i$ does not depend on $(P_i)_{i\in\N^*}$.

\begin{align*}
C_i&=
\begin{pmatrix}
\alpha_0P_{d(i-1)+1}& \alpha_1P_{d(i-1)+2} & \cdots &\alpha_{d-1}P_{di}\\
            0          & \alpha_0P_{d(i-1)+2} & \cdots &\alpha_{d-2}P_{di}\\
         \vdots        &         \ddots          & \ddots &           \vdots        \\
            0          &         \cdots          &    0   & \alpha_0P_{di} \\
\end{pmatrix}\\
\intertext{and}
D_i&=
\begin{pmatrix}
{\alpha_dP_{d(i-1)+1}}^\dagger& {\alpha_{d-1}P_{d(i-1)+1}}^\dagger & \cdots &{\alpha_1P_{d(i-1)+1}}^\dagger \\
          0        & {\alpha_dP_{d(i-1)+2}}^\dagger & \cdots & {\alpha_2P_{d(i-1)+2}}^\dagger \\
       \vdots      &       \ddots        & \ddots &          \vdots       \\
          0        &       \cdots        &    0   &{\alpha_dP_{di}}^\dagger\\
\end{pmatrix}.
\end{align*}

Let us define another channel transfer matrix $\widetilde{H}_m$ by $\widetilde{K}=1$ and for $i\in\N^*$ and $0\leq s\leq d$,
\[\widetilde{\zeta}_{i,i+s}=\alpha_s \nrm{P_{i+s}}.\]
In the same manner, we define $\widetilde{C}_i$, $\widetilde{D}_i$ and $\widetilde{\Delta_i}$. A straight forward verification shows that for $i\in\N^*$
\[C_iC_i^\dagger = \widetilde{C}_i\widetilde{C}_i^\dagger,\ C_iD_i = \widetilde{C}_i\widetilde{D}_i\textrm{ and }D_i^\dagger D_i = \widetilde{D}_i^\dagger\widetilde{D}_i.\]
Moreover, since $\Delta_i$ is a function of $C_iC_i^\dagger$, $C_iD_i$ and $D_i^\dagger D_i$, $\Delta_i=\widetilde{\Delta}_i$. However, since $\widetilde{K}=1$, we have already proved that $\widetilde{\Delta}_i$ does not depend on $(P_i)_{i\in\N^*}$, therefore, $\Delta_i$ does not depend on $(P_i)_{i\in\N^*}$.
\end{proof}

%%%%%%%%%%%%%%%%%%%%%%%%%%%%%%%%%%%%%%%%%
%%%%%%%%%%%%%%%%%%%%%%%%%%%%%%%%%%%%%%%%%
%%%%%%%%%%%%%%%%%%%%%%%%%%%%%%%%%%%%%%%%%

\section{Numerical simulations}
\label{sec:NumericalSimulations}

%%%%%%%%%%%%%%%%%%%%%%%%%%%%%%%%%%%%%%%%%
%%%%%%%%%%%%%%%%%%%%%%%%%%%%%%%%%%%%%%%%%
%%%%%%%%%%%%%%%%%%%%%%%%%%%%%%%%%%%%%%%%%

%%%%%%%%%%%%%%%%%%%%%%%%%%%%%%%%%%%%%%%%%
%%%%%%%%%%%%%%%%%%%%%%%%%%%%%%%%%%%%%%%%%
%%%%%%%%%%%%%%%%%%%%%%%%%%%%%%%%%%%%%%%%%

\subsection{Influence of the correlation}
\label{sec:InfluenceOfTheCorrelation}

%%%%%%%%%%%%%%%%%%%%%%%%%%%%%%%%%%%%%%%%%
%%%%%%%%%%%%%%%%%%%%%%%%%%%%%%%%%%%%%%%%%
%%%%%%%%%%%%%%%%%%%%%%%%%%%%%%%%%%%%%%%%%

We assume that the fading coefficients are Rayleigh distributed (real and imaginary parts are independent Gaussian random variables with zero mean and variance $1/\sqrt{2}$) and independent for different users. We are interested in the following question, which of the non-fading channel and the Rayleigh fading channel gives a higher per-cell sum-rate capacity.

In the case $d=2$, $\lambda=0.1,1$ and $K=1,2,4,10$, with all fading coefficients independent, it is shown in Subsection \ref{sec:NumericalComparaisonOfTheBounds} that the Rayleigh fading is beneficial.

In the case $d=1$ if we assume independence between the $\zeta_{i,j}$, it is known that Rayleigh fading is beneficial over non-fading channels in the high-SNR region already for $K=2$ (\cite{bidiagonal}). If we assume that for $i\in\N^*$, $\zeta_{i-1,i}=\zeta_{i,i}$, then, the sum-rate per-cell capacity is less than the one of a non-fading channel (see Subsection \ref{sec:ArtificialFading} and \cite{Somekh-Shamai-2000}). We investigate the following question: what is the maximal level of correlation between $\zeta_{i-1,i}$ and $\zeta_{i,i}$ that still provides benefit over the non-fading channel. See Appendix \ref{sec:CapacityOfTheNonFadingChannels} for the derivation of the capacity of the non-fading channels. We denote by $c$ the correlation between the real (resp. imaginary) part of $\zeta_{i,i}$ and the real (resp. imaginary) part of $\zeta_{i-1,i}$

\begin{figure}
\begin{center}
\psfrag{TTTTheoremgencorrelation}{\scriptsize Proposition \ref{bound-lyapunov}.\ref{nstepgeneral}}
\psfrag{withoutfading}{\scriptsize No fading}
\psfrag{coupelower}{\scriptsize Proposition \ref{truncation} (LB)}
\psfrag{coupeupper}{\scriptsize Proposition \ref{truncation} (UB)}
\includegraphics[scale=0.7]{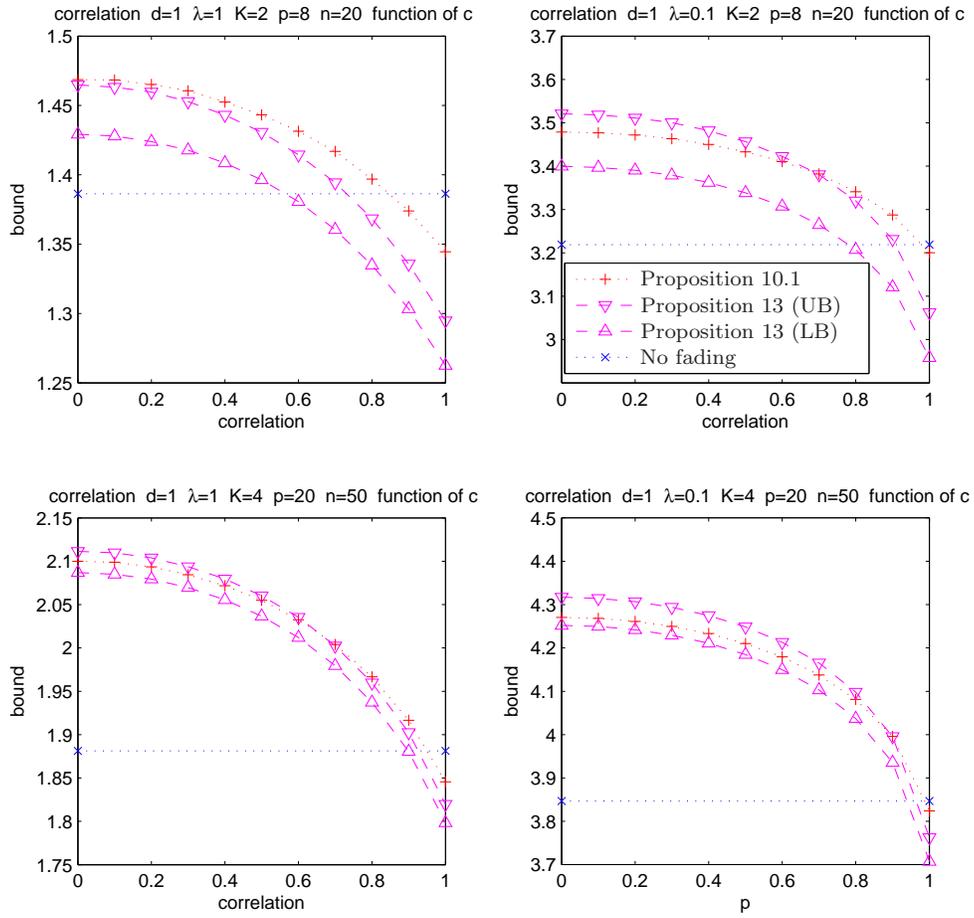}
\end{center}
\caption{Influence of the correlation on the capacity in function of the SNR and $K$. Bounds of Proposition \ref{truncation}.}
\label{fig: correlation}
\end{figure}

\begin{figure}
\begin{center}
\psfrag{TTTTheoremgencorrelation}{\scriptsize Proposition \ref{bound-lyapunov}.\ref{nstepgeneral}}
\psfrag{withoutfading}{\scriptsize No fading}
\psfrag{coupelower}{\scriptsize Proposition \ref{truncation} (LB)}
\psfrag{coupeupper}{\scriptsize Proposition \ref{truncation} (UB)}
\includegraphics[scale=0.7]{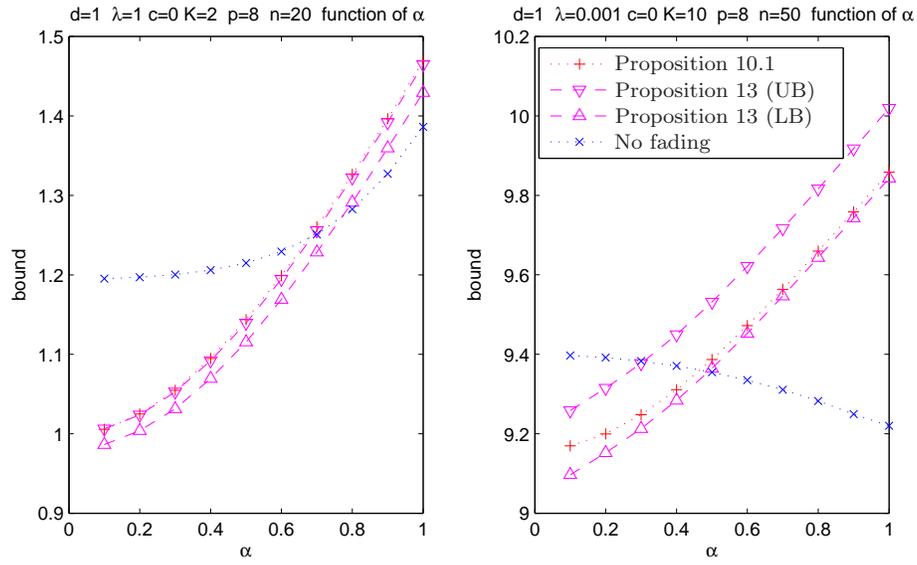}
\end{center}
\caption{Influence of the correlation on the capacity in function of $\alpha$. Bounds of Proposition \ref{truncation}.}
\label{fig: alpha_correlation}
\end{figure}

In Figure \ref{fig: correlation} we present the bounds of Proposition \ref{bound-lyapunov}.\ref{nstepgeneral} and Proposition \ref{truncation} in the special case of Rayleigh fading. In Figure \ref{fig: alpha_correlation} we present the bounds of Proposition \ref{bound-lyapunov}.\ref{nstepgeneral} and Proposition \ref{truncation} in the following special case: $\zeta_{i,i}$ is Rayleigh distributed, $\zeta_{i-1,i}$ is $\alpha\in[0,1]$ times a Rayleigh distributed random variable. In both cases, the curves are produced by Monte Carlo simulation with $10^5$ samples.

We see that even with a correlation close to 1, fading still provides an advantage over non-fading channel. Moreover, note that $K$ large, high SNR and $\alpha$ close to 1 are conditions in which the advantage of the fading is larger.

%%%%%%%%%%%%%%%%%%%%%%%%%%%%%%%%%%%%%%%%%
%%%%%%%%%%%%%%%%%%%%%%%%%%%%%%%%%%%%%%%%%
%%%%%%%%%%%%%%%%%%%%%%%%%%%%%%%%%%%%%%%%%

\subsection{The asymmetric Wyner model}
\label{sec:TheAsymmetricWynerModel}

%%%%%%%%%%%%%%%%%%%%%%%%%%%%%%%%%%%%%%%%%
%%%%%%%%%%%%%%%%%%%%%%%%%%%%%%%%%%%%%%%%%
%%%%%%%%%%%%%%%%%%%%%%%%%%%%%%%%%%%%%%%%%

With the following specification, the model studied is the Rayleigh-fading Wyner model (\cite{Wyner-94}). We take $d=2$ and the $\zeta_{i,j}$ independent with the following distributions. For $i\in\N^*$, $\zeta_{i,i+1}$ is Rayleigh distributed (real and imaginary parts are independent Gaussian random variables with zero mean and variance $1/\sqrt{2}$) and $\zeta_{i,i}$ (resp. $\zeta_{i,i+2}$) is $\alpha\in[0,1]$ times a Rayleigh distributed random variable. The asymmetric (Rayleigh-fading) Wyner model is similar to Rayleigh-fading Wyner with a slight modification. For $i\in\N^*$, $\zeta_{i,i}$ is Rayleigh distributed and $\zeta_{i,i+1}$ (resp. $\zeta_{i,i+2}$) is $\alpha$ times a Rayleigh distributed random variable. Note that in Subsection \ref{sec:CaseD2} we prove that in the asymmetric case, the power offset for $\alpha\leq0.4$ is $\gamma/\log2$.

\begin{figure}
\begin{center}
\psfrag{SSSSSSSSSSSSsymcoupeupper}{\scriptsize Symmetric (UB)}
\psfrag{symcoupelower}{\scriptsize Symmetric (LB)}
\psfrag{anticoupeupper}{\scriptsize Asymmetric (UB)}
\psfrag{anticoupelower}{\scriptsize Asymmetric (LB)}
\psfrag{symwithoutfading}{\scriptsize Symmetric non-fading}
\psfrag{antiswithoutfading}{\scriptsize Asymmetric non-fading}
\includegraphics[scale=0.6]{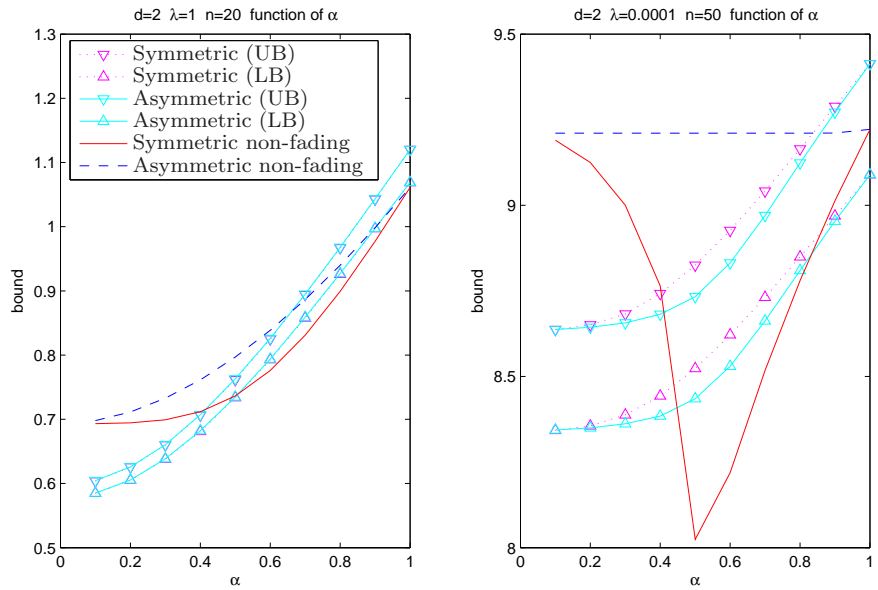}
\end{center}
\caption{Comparison of the symmetric and asymmetric channels with and without Rayleigh fading for $\lambda=1,10^{-4}$ in function of $\alpha$. Bounds of Proposition \ref{truncation}.}
\label{fig: asymetric}
\end{figure}

The two models are very similar and yet, in the non-fading case, the per-cell sum-rate capacity is notably different (see Appendix \ref{sec:CapacityOfTheNonFadingChannels} for the derivation of the capacity of the non-fading channels). In Figure \ref{fig: asymetric} we present the capacity of the two models without fading and the bounds of Proposition \ref{truncation} for the two models with Rayleigh fading. We study one case in moderate SNR ($\lambda=1$) and one case in high SNR ($\lambda=10^{-4}$). The curves are produced by Monte Carlo simulation with $10^5$ samples.

\begin{figure}
\begin{center}
\psfrag{SSSSSSSsymcoupeupper}{\scriptsize Symmetric (UB)}
\psfrag{symcoupelower}{\scriptsize Symmetric (LB)}
\psfrag{anticoupeupper}{\scriptsize Asymmetric (UB)}
\psfrag{anticoupelower}{\scriptsize Asymmetric (LB)}
\includegraphics[scale=0.6]{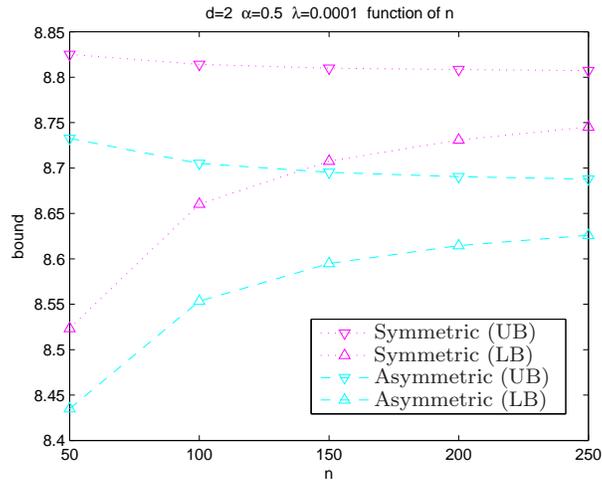}
\end{center}
\caption{Comparison of the symmetric and asymmetric channels with Rayleigh fading for $\lambda=10^{-4}$ and $\alpha=0.5$. Bounds of Proposition \ref{truncation}.}
\label{fig: asymetric2}
\end{figure}

Note that in the high-SNR region, for the non-fading channel, the per-cell sum-rate capacity is very different for symmetric and the asymmetric models, whereas the per-cell sum-rate capacities for the symmetric and asymmetric Rayleigh-fading models are very close (but not equal as shown in Figure \ref{fig: asymetric2} for $\lambda=10^{-4}$ and $\alpha=0.5$).

\begin{figure}
\begin{center}
\psfrag{SSSSSsymcoupeupper}{\scriptsize Symmetric (UB)}
\psfrag{symcoupelower}{\scriptsize Symmetric (LB)}
\psfrag{anticoupeupper}{\scriptsize Asymmetric (UB)}
\psfrag{anticoupelower}{\scriptsize Asymmetric (LB)}
\includegraphics[scale=0.7]{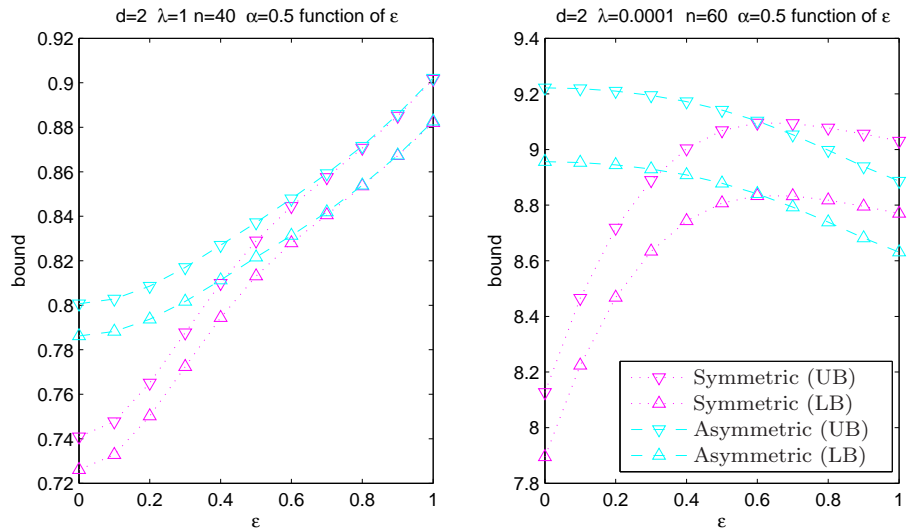}
\end{center}
\caption{Comparison of the symmetric and asymmetric channels with uniform fading for $\lambda=1,10^{-4}$ and $\alpha=0.5$ in function of $\ep$. Bounds of Proposition \ref{truncation}.}
\label{fig: asymetric_uniform}
\end{figure}

To understand better the influence of fading on the difference between the two models, we present in Figure \ref{fig: asymetric_uniform} the bounds of Proposition \ref{truncation} for the capacity of the two models (symmetric and asymmetric) with the following fading: the modulus is uniformly distributed between $1-\ep$ and $1+\ep$ and the phase is uniformly distributed between 0 and $2\ep\pi$, where $\ep$ is a parameter between 0 and 1. Note that for $\ep=0$, there is no fading and for $\ep=1$, the fading is uniformly distributed on the disc of center 0 and of radius 2. The curves are produced by Monte Carlo simulation with $10^5$ samples. We notice that the difference between the two models decreases between $\ep=0$ and $\ep=0.5$ and that in high-SNR, it increases slightly between $\ep=0.5$ and $\ep=1$.

%%%%%%%%%%%%%%%%%%%%%%%%%%%%%%%%%%%%%%%%%
%%%%%%%%%%%%%%%%%%%%%%%%%%%%%%%%%%%%%%%%%
%%%%%%%%%%%%%%%%%%%%%%%%%%%%%%%%%%%%%%%%%

\section{Concluding Remarks}
\label{sec:ConcludingRemarks}

%%%%%%%%%%%%%%%%%%%%%%%%%%%%%%%%%%%%%%%%%
%%%%%%%%%%%%%%%%%%%%%%%%%%%%%%%%%%%%%%%%%
%%%%%%%%%%%%%%%%%%%%%%%%%%%%%%%%%%%%%%%%%

In this paper, we study the per-cell sum-rate capacity of a channel communication with multiple cell processing. The main tools is a version of the Thouless formula for the strip which we prove in the article. It allows us to prove that the per-cell sum-rate capacity converges as the number of cells and antennas goes to infinity. We give several expressions of the limiting capacity in terms of Lyapunov exponents and several bounds on the per-cell sum-rate capacity.

We apply those results to several examples of communication channels and get insight on the evolution of the capacity as a function of the key parameters of the problem. In particular, in the high-SNR regime, some explicit formulas are derived.

Note that the model here applies verbatim to randomly varying intersymbol interference channels.

Some of the tools of this article can be used to derive CLT-type results on the capacity in order to study the outage-probability. Details will appear elsewhere \cite{nathanthesis}.

%%%%%%%%%%%%%%%%%%%%%%%%%%%%%%%%%%%%%%%%%
%%%%%%%%%%%%%%%%%%%%%%%%%%%%%%%%%%%%%%%%%
%%%%%%%%%%%%%%%%%%%%%%%%%%%%%%%%%%%%%%%%%

\section*{Acknowledgments}

%%%%%%%%%%%%%%%%%%%%%%%%%%%%%%%%%%%%%%%%%
%%%%%%%%%%%%%%%%%%%%%%%%%%%%%%%%%%%%%%%%%
%%%%%%%%%%%%%%%%%%%%%%%%%%%%%%%%%%%%%%%%%

We thank Oren Somekh for his help with the derivation of the capacity of the non-fading channels.

This research was partially supported by Technion Research Funds, the REMON Consortium, a grant from the Israel Science Foundation and NSF grant DMS-0503775.

\appendix
%%%%%%%%%%%%%%%%%%%%%%%%%%%%%%%%%%%%%%%%%
%%%%%%%%%%%%%%%%%%%%%%%%%%%%%%%%%%%%%%%%%
%%%%%%%%%%%%%%%%%%%%%%%%%%%%%%%%%%%%%%%%%

\subsection{Random Schr\"odinger operators techniques}
\label{sec:RandomSchrOdingerOperatorsTechniques}

%%%%%%%%%%%%%%%%%%%%%%%%%%%%%%%%%%%%%%%%%
%%%%%%%%%%%%%%%%%%%%%%%%%%%%%%%%%%%%%%%%%
%%%%%%%%%%%%%%%%%%%%%%%%%%%%%%%%%%%%%%%%%

%%%%%%%%%%%%%%%%%%%%%%%%%%%%%%%%%%%%%%%%%
%%%%%%%%%%%%%%%%%%%%%%%%%%%%%%%%%%%%%%%%%
%%%%%%%%%%%%%%%%%%%%%%%%%%%%%%%%%%%%%%%%%

\subsubsection{Lyapunov exponents theory}
\label{sec:LyapunovExponentsTheory}

%%%%%%%%%%%%%%%%%%%%%%%%%%%%%%%%%%%%%%%%%
%%%%%%%%%%%%%%%%%%%%%%%%%%%%%%%%%%%%%%%%%
%%%%%%%%%%%%%%%%%%%%%%%%%%%%%%%%%%%%%%%%%

We use the theory of product of random matrices. For a general introduction to the aspects of the theory we use here, the reader may consult \cite{carmona-lacroix}, \cite{bougerol-lacroix}, \cite{CKN}, \cite{ledrappier}, \cite{pastur-figotin} or \cite{watkins}. See appendix \ref{sec:ExteriorProduct} for the relevant background on exterior products.

\begin{thm}[Furstenberg H., Kesten H. (1960)]
\label{FK}
Consider a stationary ergodic sequence of complex random matrices $(X_i)_{i\in\N^*}$ of size $p$ and any norm on the matrices. Assume moreover that
\[\E\log^+\nrm{X_1}<\infty,\]
then a.s, $n^{-1}\log\nrm{X_n\cdots X_1}$ converges to a constant:
\[\lim_{n\rightarrow\infty}\frac{1}{n}\log\nrm{X_n\cdots X_1}\triangleq\gamma(X).\]
\end{thm}

We define $p$ constants $\gamma_1(X),\dots,\gamma_p(X)$ such that for $1\leq i\leq p$,
\[\gamma\left(\bigwedge^i X\right)=\gamma_1(X)+\cdots+\gamma_i(X).\]
\begin{prop}
\[\gamma_1(X)\geq\cdots\geq\gamma_p(X).\]
\end{prop}
The constants $\gamma_1(X)\geq\cdots\geq\gamma_p(X)$ are called the Lyapunov exponents and $\gamma(X)=\gamma_1(X)$ is called the top Lyapunov exponent.

We will also use the three following properties:
\begin{enumerate}
	\item For any sub-multiplicative norm, for $p\in\N^*$
\begin{equation}
\label{subadditive}
\gamma(X)\leq\frac{1}{p}\E\log\nrm{X_p\cdots X_1},
\end{equation}
and the limit of the RHS as $p$ goes to infinity is $\gamma(X)$.
	\item
\begin{equation}
\label{detlyapunov}
\frac{1}{p}\E\log\abs{\det X_1}\leq\gamma(X).
\end{equation}
	\item Assume that the matrices $(X_i)_{i\in\N^*}$ are i.i.d, then for all $1\leq i\leq p$, $\gamma_i(X)=\gamma_i(X^\dagger)$.
\end{enumerate}

Finally, we quote the following proposition \cite[Proposition 1]{hennion}.
\begin{prop}
\label{reductible}
Consider a stationary ergodic sequence of complex random matrices $(X_i)_{i\in\N^*}$ of size $p$ and any norm on the matrices. Assume moreover that
\[\E\log^+\nrm{X_1}<\infty.\]
Finally, assume that there exist three sequences of random matrices $(X_i^1)_{i\in\N^*}$, $(X_i^2)_{i\in\N^*}$, $(X_i^3)_{i\in\N^*}$, of respective sizes $k\times k$, $(p-k)\times k$ and $(p-k)\times(p-k)$, for $1\leq k\leq p-1$, such that almost surely, for all $i$
\[X_i=\left(\begin{array}{c|c}X_i^1&0_{k,p-k}\\\hline X_i^2&X_i^3\end{array}\right).\]
Then, $\gamma_1(X),\dots,\gamma_p(X)$ is equal up to the order to the sequence \[\gamma_1(X^1),\dots,\gamma_k(X^1),\gamma_1(X^3),\dots,\gamma_{p-k}(X^3).\]
\end{prop}

%%%%%%%%%%%%%%%%%%%%%%%%%%%%%%%%%%%%%%%%%
%%%%%%%%%%%%%%%%%%%%%%%%%%%%%%%%%%%%%%%%%
%%%%%%%%%%%%%%%%%%%%%%%%%%%%%%%%%%%%%%%%%

\subsubsection{Proof of Theorem \ref{principalK}.1}
\label{sec:ProofOfTheoremRefPrincipalK1}

%%%%%%%%%%%%%%%%%%%%%%%%%%%%%%%%%%%%%%%%%
%%%%%%%%%%%%%%%%%%%%%%%%%%%%%%%%%%%%%%%%%
%%%%%%%%%%%%%%%%%%%%%%%%%%%%%%%%%%%%%%%%%

In order to prove point 1 of Theorem \ref{principalK}, we first prove a slightly more general lemma.

\begin{lem}
\label{thouless-strip}
Assume (H\ref{ergodicity}), (H\ref{moments}) and (H\ref{frontier}).
For all $\lambda\in\C$ such that $\lambda\notin\R^-$, almost surely,
\[\frac{1}{dn}\log\abs{\det G_{dn}}\tendvers\frac{1}{d}{\E\log\abs{\det(C_2D_2)}}+\frac{1}{d}\gamma\left(N\right).\]
\end{lem}

\begin{proof}
For $i\in\N^*$, set $B_i=C_iC_i^\dagger+D_{i+1}^\dagger D_{i+1}+\lambda\Id_d$ and $A_i=C_i D_i$. Note that the eigenvalues of $G_{dn}$ are bounded away from zero. To compute $\log\abs{\det G_{dn}}$, we write the following decomposition: $G_{dn}U_{dn}=L_{dn}$, where $U_{dn}$ is the upper triangular by block matrix
\[U_{dn}=
\left(\begin{array}{c|c|c|c}
   X_1  &   X_1  & \cdots &   X_1  \\\hline
    0_d   &   X_2  & \cdots &   X_2  \\\hline
 \vdots &   \ddots  & \ddots & \vdots \\\hline
    0_d   & \cdots &    0_d   &   X_n  \\
\end{array}\right),\]
the $X_i$ are $d\times d$ matrices such that $X_0=0_d$, $X_1=\Id_d$, and for $i\geq 1$,
\begin{equation}
\label{recursion}
A_{i} X_{i-1}+B_i X_i+A_{i+1}^\dagger X_{i+1}=0_d.
\end{equation}

$L_{dn}$ is the lower triangular by block matrix
\[\left(\begin{array}{c|c|c|c}
  -A_2^\dagger X_2  &   0_d  & \cdots &   0_d  \\\hline
    A_2X_1   &   -A_3^\dagger X_3  & \ddots &  \vdots  \\\hline
   0_d   &   \ddots  & \ddots &    0_d   \\\hline
    0_d   &    0_d    &  	A_{n}X_{n-1}   &   -A_{n+1}^\dagger X_{n+1}  \\
\end{array}\right),\]

That decomposition allows us to write $\log\abs{\det G_{dn}}$ as a determinant by block,
\begin{align*}
\log\abs{\det G_{dn}}&=\sum_{i=1}^n\log\abs{\det A_{i+1}}+\sum_{i=1}^n\log\abs{\det X_{i+1}}-\sum_{i=1}^n\log\abs{\det X_i}\\
&=\sum_{i=1}^n\log\abs{\det A_{i+1}}+\log\abs{\det X_{n+1}}.
\end{align*}
Therefore
\begin{equation}
\label{narula_derivation}
\frac{1}{dn}\log\abs{\det G_{dn}}=\frac{1}{dn}\sum_{i=1}^n\log\abs{\det A_{i+1}}+\frac{1}{dn}\log\abs{\det X_{n+1}}.
\end{equation}
$\frac{1}{n}\sum_{i=1}^n\log\abs{\det A_{i+1}}$ converges by ergodicity toward $\E\log\abs{\det A_2}$. Note that the choice of $A_{n+1}$ is arbitrary, indeed, if we take another value, say $\widetilde{A}_{n+1}$, then $\widetilde{A}^\dagger_{n+1}\widetilde{X}_{n+1}=A_{n+1}^\dagger X_{n+1}$ and (\ref{narula_derivation}) stays unchanged.

We emphasize that the derivation of (\ref{narula_derivation}) is inspired by Narula's thesis (\cite{Narula-1997}). 

The $X_i$ are defined by (\ref{recursion}). We can reformulate it in the following way. Set $V_i=\left(\begin{array}{c}X_{i-1}\\\hline X_i\end{array}\right)$, then (\ref{recursion}) is equivalent to $V_{i+1}=M_iV_i$ and moreover,
\[X_{n+1}=\left(\begin{array}{c|c}0_d&\Id_d\end{array}\right)M_n\cdots M_1\left(\begin{array}{c}0_d\\\hline\Id_d\end{array}\right).\]

Denote $f=\bigwedge\left(\begin{array}{c}0_d\\\hline\Id_d\end{array}\right)$. For the relevant background on exterior products, see Appendix \ref{sec:ExteriorProduct}. We get
\[N_{n}N_{n-1}\dots N_1\left(v^1_1\wedge\dots\wedge v^d_1\right).\]
However, $v^1_1\wedge\dots\wedge v^d_1=f$. Therefore,
\begin{equation}
\label{M}
\begin{split}
\det X_{n+1}&=\bigwedge^d X_{n+1}=\bigwedge^d\left(\left(\begin{array}{c|c}0_d&\Id_d\end{array}\right)M_n\cdots M_1\left(\begin{array}{c}0_d\\\hline\Id_d\end{array}\right)\right)\\
&=f^\dagger N_n N_{n-1}\dots N_1 f.
\end{split}
\end{equation}

Taking the canonical basis of $\bigwedge^d\C^{2d}$, $f$ is the last vector of the basis and $\det X_{n+1}$ grows like the bottom-right coefficient of the product of the $N_i$, therefore, its growth rate is bounded above by the Lyapunov exponent of the $N_i$.
\begin{equation}
\label{upper_bound}
\limsup_{n\rightarrow\infty}\frac{1}{n}\E\log\abs{\det X_{n+1}}\leq\gamma(N).
\end{equation}
Using (\ref{narula_derivation}), it is enough to prove the opposite inequality to conclude the proof. The end of the proof is inspired by \cite{Craig-Simon}.

\begin{lem}
If there exist a basis of $\bigwedge^d\C^{2d}$, say $(g_i)_{i\in I}$, such that for all $i,j\in I$, almost surely,
\begin{equation}
\label{condition}
\liminf_{n\rightarrow\infty}\frac{1}{n}\log\abs{g_j^\dagger N_{n}\cdots N_1g_i}\leq\liminf_{n\rightarrow\infty}\frac{1}{n}\log\abs{\det X_{n+1}},
\end{equation}
then, almost surely,
\[\gamma(N)\leq\liminf_{n\rightarrow\infty}\frac{1}{n}\log\abs{\det X_{n+1}}.\]
\end{lem}
Let us first prove the lemma.

\begin{proof}
For any finite basis $S_1$ and $S_2$ in a vector space, we have for all $A$
\[\sup_{\alpha\in S_1,\beta\in S_2}\abs{\alpha^\dagger A\beta}\geq c\nrm{A}\]
for some universal $c$. Thus, (\ref{condition}) shows that, almost surely,
\[\gamma(N)\leq\liminf_{n\rightarrow\infty}\frac{1}{n}\log\abs{\det X_{n+1}}.\]
\end{proof}

To finish the proof of Lemma \ref{thouless-strip}, we denote by $\{e_1,\dots,e_{2d}\}$ the canonical basis of $\C^{2d}$ and we apply the lemma with the following spanning system of $\bigwedge^d\C^{2d}$
\[S\triangleq\{(e_1+\widetilde{e}_1)\wedge\cdots\wedge(e_d+\widetilde{e}_d)\ ,\ \widetilde{e}_1,\dots,\widetilde{e}_d\in\vct(e_{d+1},\dots,e_{2d})\}.\]
For a choice of $e^\#_1,\dots,e^\#_d$, such that for $1\leq j\leq d$, $e^\#_j=\sum_{i=1}^d\alpha_{i,j}e_{d+i}$, we define $E^\#$ the $d\times d$ matrix of the $\alpha_{i,j}$. We get
\[g_1\triangleq(e_1+e^\#_1)\wedge\cdots\wedge(e_d+e^\#_d)=\bigwedge^d\left(\begin{array}{c}Id_d\\\hline E^\#\end{array}\right).\]
In the same way, for a choice of $e^b_1\dots,e^b_d\in\vct(e_{d+1},\dots,e_{2d})$, we define $E^b$, a $d\times d$ matrix, such that
\[g_2\triangleq(e_1+e^b_1)\wedge\cdots\wedge(e_d+e^b_d)=\bigwedge^d\left(\begin{array}{c}Id_d\\\hline E^b\end{array}\right).\]

We define two new sequences $(\widetilde{A}_i)$ and $(\widetilde{B}_i)$ such that
\begin{itemize}
	\item For $2\leq i\leq n-1$, $\widetilde{B}_i=B_i$,
	\item For $1\leq i\leq n$, $\widetilde{A}_i=A_i$,
	\item $\widetilde{B}_1=-A_2^\dagger E^\#$,
	\item $\widetilde{B}_n=A_n \left(E^b\right)^\dagger$,
	\item $\widetilde{A}_{n+1}=-A_n^\dagger$.
\end{itemize}
We also define $\widetilde{G}_{dn}$, $\widetilde{M}_i$, $\widetilde{N}_i$ and $\widetilde{X}_i$ using $\left(\widetilde{A}_i\right)_{i\in\N^*}$ and $\left(\widetilde{B}_i\right)_{i\in\N^*}$. Then,
\begin{align*}
f^\dagger \widetilde{N}_n \widetilde{N}_{n-1}\dots \widetilde{N}_1 f&=
f^\dagger
\bigwedge^d\left(\begin{array}{c|c}0_d&\Id_d\\\hline\Id_d&\left(E^b\right)^\dagger\end{array}\right)
\widetilde{N}_{n-1}\dots \widetilde{N}_2
\bigwedge^d\left(\begin{array}{c|c}0_d&\Id_d\\\hline-T_1T_2^{-1\dagger}&E^\#\end{array}\right) f\\
&=g_2^\dagger\widetilde{N}_{n-1}\dots \widetilde{N}_2 g_1.
\end{align*}
Therefore, to prove the condition (\ref{condition}), it is enough to prove that, almost surely,
\[\limsup_{n\rightarrow\infty}\left(\frac{1}{n}\log\abs{\det\widetilde{G}_{dn}}-\frac{1}{n}\log\abs{\det{G_{dn}}}\right)\leq 0.\]
We now use perturbation theory techniques. Indeed, we denote by $\rho$ the spectral radius of a matrix, i.e. its largest eigenvalue in absolute value. Recall that for a matrix $S$, $\rho(S)\leq\sqrt{\rho(SS^\dagger)}$ and that $\sqrt{\rho(SS^\dagger)}$ is a sub-multiplicative norm. As a consequence, for positive Hermitian matrices, the spectral radius is sub-multiplicative. Moreover, we denote by $\nrm{\cdot}_F$ the Fr\"obenius norm. Recall that $\sqrt{\rho(SS^\dagger)}\leq\nrm{S}_F$.  We will also use the fact that the eigenvalues of $G_{dn}$ are bounded away from 0 by $\mu=\lambda$ if $\lambda>0$ or $\mu=\abs{\Im\lambda}$ if $\lambda\notin\R$.
Moreover, we define $U_{dn}=\widetilde{G}_{dn}-G_{dn}$, which has rank less than or equal to $2d$.
\begin{align*}
\frac{1}{n}\log\abs{\det\widetilde{G}_{dn}}-\frac{1}{n}\log\abs{\det{G_{dn}}}&=
\frac{1}{n}\log\abs{\det(G_{dn}+U_{dn})}-\frac{1}{n}\log\abs{\det{G_{dn}}}\\
&=\frac{1}{n}\log\abs{\det(\Id_{dn}+G_{dn}^{-1}U_{dn})}.
\end{align*}
$G_{dn}^{-1}U_{dn}$ has rank at most 2d, therefore,
\begin{align*}
\frac{1}{n}\log\abs{\det\widetilde{G}_{dn}}-\frac{1}{n}\log\abs{\det{G_{dn}}}&\leq
\frac{2d}{n}\log\abs{1+\rho(G_{dn}^{-1}U_{dn})}\\
&\leq\frac{2d}{n}\log\abs{1+\sqrt{\rho(G_{dn}^{-1}U_{dn}U_{dn}^\dagger G_{dn}^{-1\dagger})}}\\
&\leq\frac{2d}{n}\log\abs{1+\sqrt{\rho(G_{dn}^{-1\dagger}G_{dn}^{-1})}\sqrt{\rho(U_{dn}U_{dn}^\dagger)}}\\
&\leq\frac{2d}{n}\log\abs{1+\frac{1}{\mu}\sqrt{\rho(U_{dn}U_{dn}^\dagger)}}\\
&\leq\frac{2d}{n}\log\abs{1+\frac{1}{\mu}\nrm{U_{dn}}_F}.
\end{align*}
Moreover,
\[\nrm{U_{dn}}_F^2=\nrm{S_1+\lambda\Id_d+T_2^\dagger E^\#}_F^2+\nrm{S_n+\lambda\Id_d+T_n^\dagger E^b}_F^2,\]
hence, with the integrability condition, $\sup_n\E\log\abs{1+\frac{1}{\mu}\nrm{U_{dn}}_F}^{1+\ep}<\infty$. By Tchebicheff inequality, for a given $\eta>0$,
\[\P\left(\frac{1}{n}\log\abs{1+\frac{1}{\mu}\nrm{U_{dn}}_F}>\eta\right)\leq\frac{\sup_n\E\log\abs{1+\frac{1}{\mu}\nrm{U_{dn}}_F}^{1+\ep}}{(\eta n)^{1+\ep}}.\]
The RHS is a summable series, therefore, by Borel-Cantelli Lemma, almost surely,
\[\limsup_{n\rightarrow\infty}\left(\frac{1}{n}\log\abs{\det\widetilde{G}_{dn}}-\frac{1}{n}\log\abs{\det{G_{dn}}}\right)\leq 0.\]
This finishes the proof of Lemma \ref{thouless-strip}.
\end{proof}

\begin{rmk}
\label{perturbation}
By the same kind of perturbation theory techniques, we can show that in order to study the limit in $m$ of $Cap_m(\rho)$, it is enough to study the sequence every $d$ steps.
\end{rmk}

For a hermitian matrix $h$ whose ordered eigenvalues are $\alpha_1,\dots,\alpha_n$, we denote by the \emph{spectral distribution} of $h$, the measure
\[\frac{1}{n}\sum_{i=1}^n\delta_{\alpha_i},\]
where $\delta_x$ is a Dirac measure at $x$.

The following technical lemma will be used several times to prove domination properties.
\begin{lem}
\label{domination}
Denote by $\mu_n$ the spectral distribution of $H_{dn}H_{dn}^\dagger$. Consider the following diagonal by blocks matrix:
\[F_{dn}\triangleq\left(\begin{array}{c|c|c|c}
  2B_1  &   0_d  & \cdots &   0_d  \\\hline
    0_d   &   2B_2  & \ddots &  \vdots  \\\hline
   \vdots   &   \ddots  & \ddots &    0_d   \\\hline
    0_d   &    \cdots    &  0_d   &   2B_n  \\
\end{array}\right),\]
and denote by $\tilde{\mu}_n$ its spectral distribution. Then, for any non-decreasing function $f$,
\[\int f d\mu_n\leq\int f d\tilde{\mu}_n.\]
\end{lem}

\begin{proof}
Denote
\[\tilde{H}_{dn}=
\left(\begin{array}{c|c|c|c|c}
   C_1  & -D_2^\dagger &      0_{d,dK}      &  \cdots &      0_{d,dK}      \\\hline
    0_{d,dK}   &     C_2     & -D_3^\dagger &  \ddots &   \vdots    \\\hline
 \vdots &   \ddots    &   \ddots    &  \ddots &      0_{d,dK}      \\\hline
    0_{d,dK}   &   \cdots    &      0_{d,dK}      & C_n & -D_{n+1}^\dagger \\
\end{array}\right),\]
then $F_{dn}=H_{dn}H_{dn}^\dagger+\tilde{H}_{dn}\tilde{H}_{dn}^\dagger$. Since $\tilde{H}_{dn}\tilde{H}_{dn}^\dagger$ is a non-negative Hermitian matrix, by Weyl's inequalities, for all $1\leq i\leq dn$, the $i$-th eigenvalue of $H_{dn}H_{dn}^\dagger$ is less or equal than the $i$-th eigenvalue of $F_{dn}$.
\end{proof}
First note that $(1/d){\E\log\abs{\det(C_2D_2)}}=\E_{\pi}\log\abs{\zeta_0\zeta_d^\dagger}$. From Lemma \ref{domination}, we deduce that for $\lambda>0$,
\[\frac{1}{dn}\log\abs{\det G_{dn}}\leq\log2+\frac{1}{n}\sum_{i=1}^n\log\abs{\det B_i}.\]
Therefore by (H\ref{moments}) and Hadamard's inequality, $(1/dn)\log\abs{\det G_{dn}}$ is a uniformly integrable sequence and the almost sure convergence of Lemma \ref{thouless-strip} implies point 1 of Theorem \ref{principalK}.

%%%%%%%%%%%%%%%%%%%%%%%%%%%%%%%%%%%%%%%%%
%%%%%%%%%%%%%%%%%%%%%%%%%%%%%%%%%%%%%%%%%
%%%%%%%%%%%%%%%%%%%%%%%%%%%%%%%%%%%%%%%%%

\subsubsection{Proof of Theorem \ref{principalK}.2}
\label{sec:ProofOfTheoremRefPrincipalK2}

%%%%%%%%%%%%%%%%%%%%%%%%%%%%%%%%%%%%%%%%%
%%%%%%%%%%%%%%%%%%%%%%%%%%%%%%%%%%%%%%%%%
%%%%%%%%%%%%%%%%%%%%%%%%%%%%%%%%%%%%%%%%%

We begin by a few notations. For $\lambda\notin\R$, set
\[f(\lambda)=\lim_{n\rightarrow\infty}\frac{1}{dn}\log\abs{\det\left(H_{dn}H_{dn}^\dagger+\lambda\Id_{dn}\right)},\]
which exists by Lemma \ref{thouless-strip}. The existence of the weak limit of $\mu_n$ and the fact that it is non random is a classical fact of the random Schr\"odinger operators theory, see for example \cite[Theorem 4.4]{pastur-figotin}. For $\lambda\in\C$, we set (if it exists)
\[g(\lambda)=\int\log\abs{x+\lambda}d\mu(x).\]
We emphasize that since $\log$ is not a bounded function, we cannot directly deduce from Lemma \ref{thouless-strip} and the weak convergence of the $\mu_n$ to $\mu$ that for $\lambda\notin\R$, $f(\lambda)=g(\lambda)$.

Finally, for $\lambda\in\C$, define
\[h(\lambda)=\E_{\pi}\log\abs{\zeta_0\zeta_d^\dagger}+\frac{1}{d}\gamma\left(N\right).\]
The following lemma is a generalization of the Thouless formula for the strip proved in \cite{Craig-Simon}.
\begin{lem}
\label{schthouless}
Assume (H\ref{ergodicity}), (H\ref{moments}) and (H\ref{frontier}).
For all $\lambda\in\C$, $g(\lambda)=h(\lambda)$.
\end{lem}
The proof of this result is done in the frame of channel transfer matrices but one does not need to assume that the $A_i$ and the $B_i$ are upper triangular by blocks, one just need instead of (H\ref{frontier}) the hypothesis that almost surely, $A_i B_i$ is invertible.

\begin{proof}
The proof goes along the following lines, we first prove that for $\lambda\notin\R$, $g(\lambda)$ exists and equals to $h(\lambda)$, then, following \cite{craig-simon2} we argue that $g$ and $h$ are two subharmonic functions equal everywhere except a set of 0 measure, therefore they are equal everywhere.

\emph{Step 1: }
Let us first prove that for $\lambda\notin\R$, $g(\lambda)$ is well defined. $\log\abs{x+\lambda}$ is bounded away from $-\infty$, therefore, $g(\lambda)$ exists although it may be $\infty$. For $R\geq0$, let us denote by $\log_R$ the function $t\rightarrow\log(t)\wedge R$. By monotone convergence, it is enough to prove that $\int\log_R\abs{x+\lambda}d\mu(x)$ is bounded uniformly in $R$. Since $x\rightarrow\log_R\abs{x+\lambda}$ is a bounded continuous function,
\[\int\log_R\abs{x+\lambda}d\mu(x)=\lim_{n\rightarrow\infty}\int\log_R\abs{x+\lambda}d\mu_n(x).\]
By Lemma \ref{domination}, and using that  $\int\log_R\abs{x+\lambda}d\mu_n(x)\leq\int\log\abs{x+\lambda}d\mu_n(x)$,
\[\lim_{n\rightarrow\infty}\int\log_R\abs{x+\lambda}d\mu_n(x)\leq\lim_{n\rightarrow\infty}\int\log\abs{x+\lambda}d\tilde{\mu}_n(x)=\E\log\abs{\det B_1}<\infty,\]
where the last inequality comes from (H\ref{moments}) and Hadamard's inequality. Finally, we get that for $\lambda\notin\R$,
\[g(\lambda)\leq\E\log\abs{\det B_1}<\infty.\]

\emph{Step 2: }
Let us prove that for $\lambda\notin\R$, $f(\lambda)=g(\lambda)$. Applying Lemma \ref{domination} one shows that for $\lambda\notin\C$, the sequence $\left(\int\log\abs{x+\lambda}d\mu_n(x)\right)_{n\in\N^*}$ is uniformly integrable and therefore,
\[f(\lambda)=\lim_{n\rightarrow\infty}\E\int\log\abs{x+\lambda}d\mu_n(x).\]
By Lemma \ref{domination}, for $R\geq 0$,
\begin{align*}
\E\int_{x\geq R}\log\abs{x+\lambda}d\mu_n(x)&\leq\E\int_{x\geq R}\log\abs{x+\lambda}d\tilde{\mu}_n(x)\\
&=\E\int_{x\geq R}\log\abs{x+\lambda}d\tilde{\mu}_1(x).
\end{align*}
Therefore, for $n\in\N^*$ and $R\geq0$,
\begin{align*}
&\abs{\E\int\log\abs{x+\lambda}d\mu_n(x)-g(\lambda)}\leq\abs{\E\int_{x\geq R}\log\abs{x+\lambda}d\tilde{\mu}_1(x)}\\
&\qquad\qquad\qquad\qquad+\abs{\E\int\log_R\abs{x+\lambda}d\mu_n(x)-\E\int\log_R\abs{x+\lambda}d\mu(x)}\\
&\qquad\qquad\qquad\qquad+\abs{\E\int_{x\geq R}\log\abs{x+\lambda}d\mu(x)}.
\end{align*}
We first fix $R\geq 0$ such that the first and the third terms are arbitrary small and then, by weak convergence, the second term goes to 0 as $n$ goes to infinity. Therefore, for $\lambda\notin\R$, $f(\lambda)=g(\lambda)$ and by Lemma \ref{thouless-strip}, $g(\lambda)=h(\lambda)$.

\emph{Step 3: }
Let us prove that $g$ and $h$ are subharmonic on $\C$. See \cite{craig-simon2} for the relevant definitions. Since for $i\in\N^*$, $N_i(\lambda)$ is an entire function of $\lambda$, $h$ is subharmonic (\cite{craig-simon2}).

Let us prove that $g$ is subharmonic. For $R\geq0$, set
\[g_R(\lambda)\triangleq\int\left(\log\abs{x+\lambda}\vee -R\right)d\mu(x).\]
By Lemma \ref{domination}, $g_R$ a continuous function. As $R$ goes to infinity, $g_R$ is a decreasing sequence of functions converging point wise to $g$, therefore, $g$ is subharmonic.

The functions $g$ and $h$ are subharmonic on $\C$ and equal on $\C-\R$, therefore, $g$ and $h$ are equal on $\C$.
\end{proof}

To finish the proof of point 2 of Theorem \ref{principalK}, let us prove that $h(\lambda)$ converges to $h(0)$ when $\lambda$ goes to 0 in $\R^+$. Note that $\E\log\abs{\det M_1}=0$, therefore, $\E\log\abs{\det N_1}=0$. By (\ref{detlyapunov}), $h(0)\geq\E_{\pi}\log\abs{\zeta_0\zeta_d^\dagger}$, therefore, using Lemma \ref{domination} and the fact that for $\lambda,x\in\R^+$, $\log\abs{x+\lambda}\geq\log\abs{x}$, we get the desired result.

%%%%%%%%%%%%%%%%%%%%%%%%%%%%%%%%%%%%%%%%%
%%%%%%%%%%%%%%%%%%%%%%%%%%%%%%%%%%%%%%%%%
%%%%%%%%%%%%%%%%%%%%%%%%%%%%%%%%%%%%%%%%%

\subsection{Other proofs}
\label{sec:OtherProofs}

%%%%%%%%%%%%%%%%%%%%%%%%%%%%%%%%%%%%%%%%%
%%%%%%%%%%%%%%%%%%%%%%%%%%%%%%%%%%%%%%%%%
%%%%%%%%%%%%%%%%%%%%%%%%%%%%%%%%%%%%%%%%%

%%%%%%%%%%%%%%%%%%%%%%%%%%%%%%%%%%%%%%%%%
%%%%%%%%%%%%%%%%%%%%%%%%%%%%%%%%%%%%%%%%%
%%%%%%%%%%%%%%%%%%%%%%%%%%%%%%%%%%%%%%%%%

\subsubsection{Proof of Proposition \ref{thouless-like}}
\label{sec:ProofOfPropositionRefThoulessLike}

%%%%%%%%%%%%%%%%%%%%%%%%%%%%%%%%%%%%%%%%%
%%%%%%%%%%%%%%%%%%%%%%%%%%%%%%%%%%%%%%%%%
%%%%%%%%%%%%%%%%%%%%%%%%%%%%%%%%%%%%%%%%%

We use the notation of Subsection \ref{sec:ProofOfTheoremRefPrincipalK1}. We define $x^i_j$ for $i\in\N$ and $1\leq j\leq d$ such that the the element at the position $(s,t)$ of $X_i$ is $x^{(i-1)d+s}_t$. Recall that $G_{dn}U_{dn}=L_{dn}$. Therefore, for a given $j$ such that $1\leq j\leq d$, we get the following characterization of the sequence $(x^i_j)_i$. $x^i_j=0$ for $-d+1\leq i\leq 0$, $x^{i}_j=\delta_{i,j}$ for $1\leq i\leq d$ and for $i\geq d+1$,
\begin{equation}
\sum_{l=i-d}^{i+d}\widetilde{\zeta}_{i,l} x_j^l=0.
\end{equation}
Therefore,
\[\begin{pmatrix}x^{i-d+1}_j\\\vdots\\x^{i+d}_j\end{pmatrix}=\mathfrak{m}_i\begin{pmatrix}x^{i-d}_j\\\vdots\\x^{i+d-1}_j\end{pmatrix}.\]
Moreover,
\begin{align*}
V_{i+1}&=
\begin{pmatrix}
x_1^{(i-1)d+1}&\cdots&x_d^{(i-1)d+1}\\
\vdots& &\vdots\\
x_1^{(i+1)d}&\cdots&x_d^{(i+1)d}\\
\end{pmatrix}\\
&=\mathfrak{m}_{id}\cdots\mathfrak{m}_{(i-1)d+1}
\begin{pmatrix}
x_1^{(i-2)d+1}&\cdots&x_d^{(i-2)d+1}\\
\vdots& &\vdots\\
x_1^{id}&\cdots&x_d^{id}\\
\end{pmatrix}\\
&=\mathfrak{m}_{id}\cdots\mathfrak{m}_{(i-1)d+1}V_i.
\end{align*}
Therefore, together with \ref{recursion}, it proves Proposition \ref{thouless-like}.

%%%%%%%%%%%%%%%%%%%%%%%%%%%%%%%%%%%%%%%%%
%%%%%%%%%%%%%%%%%%%%%%%%%%%%%%%%%%%%%%%%%
%%%%%%%%%%%%%%%%%%%%%%%%%%%%%%%%%%%%%%%%%

\subsubsection{Proof of Proposition \ref{principalKindep}.2}
\label{sec:ProofOfPropositionRefPrincipalKindep2}

%%%%%%%%%%%%%%%%%%%%%%%%%%%%%%%%%%%%%%%%%
%%%%%%%%%%%%%%%%%%%%%%%%%%%%%%%%%%%%%%%%%
%%%%%%%%%%%%%%%%%%%%%%%%%%%%%%%%%%%%%%%%%

In order to prove point 2 of Proposition \ref{principalKindep}, we first prove the following lemma:

\begin{lem}
\label{decomposition}
For all $i\geq d+1$, there exist matrices $p^s_1(i)$, $p^s_2(i)$ for $1\leq s\leq d$ and $\delta^s(i)$ for $1\leq s\leq d+1$ such that $\delta^1(i)=\mu(i)$, $\delta^s(i)=\underline{\delta}^s(\zeta^i,\dots,\zeta^{i+d-s+1})$, $p^s_1(i)=\underline{p}^s_1(\zeta^i,\dots,\zeta^{i+d-s})$ and $p^s_2(i)=\underline{p}^s_2(\zeta^{i+d-s})$, where $\underline{\delta}^s$, $\underline{p}^s_1$ and $\underline{p}^s_2$ are deterministic functions. We have moreover the two relationships
\begin{equation}
\label{star1}
\delta^s(i)=p^s_2(i+1)p^s_1(i).
\end{equation}
\begin{equation}
\label{star2}
\delta^{s+1}(i)=p^s_1(i)p^s_2(i).
\end{equation}
Finally, for $i\geq d+1$, $\delta_i=\delta^{d+1}(i)$
\end{lem}

\begin{proof}
For $i\geq d+1$ and $1\leq s\leq d$, define
\begin{itemize}
	\item for $s\leq l\leq 2d$,
				\[a^{i,s}_l=-\lambda\1_{(l=d+s)}-\sum_{t=(l-s-d)\vee0}^{(l-s)\wedge(d-s)}\zeta_{i,i+t}\zeta_{i+l-d-s,i+t}^\dagger,\]
	\item for $1\leq l\leq d$, $\alpha^{i-s}_l=-\zeta_{i-s+l-1,i+d-s}^\dagger/\zeta_{i+d-s,i+d-s}^\dagger$,
	\item for $1\leq l\leq s$, $b^{i-s}_l= -\zeta_{i-s+l,i+d-s}/\zeta_{i-s,i+d-s}$,
	\item $\beta^{i-s=}1/\zeta_{i-s,i+d-s}\zeta_{i+d-s,i+d-s}^\dagger$.
\end{itemize}
Then
\setlength{\arraycolsep}{0pt}
\[p_1^s(i)=\left(
\begin{array}{c}
	\setlength{\arraycolsep}{5pt}
	\begin{array}{c|c}
		 & \\
		\Id_{s-1}&0_{s-1,2d-s+1}\\
		 & \\\hline
		0_{1,s-1}&
		\begin{array}{c c c}
			a^{i,s}_{s}&\cdots&a^{i,s}_{2d}
		\end{array}\\
	\end{array}\\\hline
	\setlength{\arraycolsep}{5pt}
	\begin{array}{c|ccc}
		 & & & \\
		0_{2d-s,s}& &\Id_{2d-s}& \\ 
		 & & &
	\end{array}
\end{array}\right)
\]
and

\setlength{\arraycolsep}{0pt}
\[p_2^s(i)=\left(
\begin{array}{c|c}
	\setlength{\arraycolsep}{5pt}
	\begin{array}{c}
		 b^{i-s}_1\\
		 \vdots\\
		 b^{i-s}_{s-1} \\\hline
		 \\ \\
		 0_{2d-s,1}\\
		 \\ \\
	 \end{array}&
	 \begin{array}{ccc}
	 	 & & \\
	 	 & \Id_{2d-1}& \\
	 	 & & \\
	 \end{array}\\\hline
	 \beta^{i-s}&
	 \setlength{\arraycolsep}{5pt}
	 \begin{array}{ccccc|ccc}
	 	 & &0_{1,d-1}& & &\alpha^{i-s}_1&\cdots&\alpha^{i-s}_d
	 \end{array}
\end{array}\right).\]

Finally, for $2\leq s\leq d$,

\[\delta^s(i)=\left(
\begin{array}{c|c}
	\setlength{\arraycolsep}{5pt}
	\begin{array}{c}
		b^{i-s+1}_1\\
		\vdots\\
		b^{i-s+1}_{s-1}\\\hline
		 \\
		0_{2d-s,1}\\
		 \\\hline
		\beta^{i-s+1}
	\end{array}&
	\begin{array}{@{\hspace{2pt}}c@{\hspace{2pt}}|c}
		\\
		\Id_{s-2}&0_{s-2,2d-s+1}\\
		\\\hline
		0_{2d-s+2,s-2}&
		\begin{array}{c}
			\setlength{\arraycolsep}{5pt}
			\begin{array}{ccc}
				a^{i,s}_{s}&\cdots\cdots\cdots\cdots\cdots\cdots&a^{i,s}_{2d}\\
			\end{array}\\\hline
			\setlength{\arraycolsep}{1pt}
			\begin{array}{c|@{\hspace{2cm}}c@{\hspace{2cm}}}
			  \\
				0_{2d-s,1}&\Id_{2d-s}\\
				\\
			\end{array}\\\hline
			\setlength{\arraycolsep}{5pt}
			\begin{array}{ccc|ccc}
				& 0_{1,d-s+1}& &\alpha^{i-s+1}_1&\cdots&\alpha^{i-s+1}_d
			\end{array}
		\end{array}
	\end{array}
\end{array}\right).\]

\setlength{\arraycolsep}{5pt}

A (straight forward yet tedious) verification shows that (\ref{star1}) and (\ref{star2}) are satisfied.

\end{proof}
Note that in the proof, we make a choice of particular $\underline{p}^s_1$, $\underline{p}^s_2$ and $\underline{\delta}^s$. Point 2 of Proposition \ref{principalKindep} is a direct consequence of the following lemma

\begin{lem}
\label{division2}
For all $i\in\N^*$,
\[\Delta_i=\delta_{id}\cdots\delta_{(i-1)d+1}.\]
Therefore,
\[\Xi_i=\xi_{id}\cdots\xi_{(i-1)d+1}.\]
\end{lem}

\begin{proof}
With the matrices of Lemma \ref{decomposition}, we can transform the product of the $\mu_i$ using alternatively (\ref{star1}) and (\ref{star2}).
\begin{align*}
M_i&=P_2(i+1)P_1(i)\\
&=P_2(i+1)\Delta(i)\left(P_2(i)\right)^{-1},
\end{align*}
\begin{align*}
&\mu_{id}\cdots\mu_{(i-1)d+1}\\
&=\delta^1(id)\cdots\delta^1((i-1)d+1)\\
&=p^1_2(id+1)p^1_1(id)p^1_2(id)p^1_1(id-1)\cdots p^1_2((i-1)d+2)p^1_1((i-1)d+1)\\
&=p^1_2(id+1)\delta^2(id)\cdots\delta^2((i-1)d+1)\left(p^1_2((i-1)d+1)\right)^{-1}\\
&=p^1_2(id+1)\cdots p^d_2(id+1)\delta^{d+1}(id)\cdots\delta^{d+1}((i-1)d+1)\left[p^1_2((i-1)d+1)\cdots p^d_2((i-1)d+1)\right]^{-1},\\
\end{align*}
where the last equality is proved by induction.

Therefore
\begin{align*}
&P_2(i+1)\Delta(i)\left(P_2(i)\right)^{-1}=\\
&p^1_2(id+1)\cdots p^d_2((id+1))\delta_{id}\cdots\delta_{(i-1)d+1}\left(p^1_2((i-1)d+1)\cdots p^d_2((i-1)d+1)\right)^{-1}
\end{align*}
and
\begin{equation}
\label{eq1}
\begin{split}
&\left[p^1_2(id+1)\cdots p^d_2((id+1))\right]^{-1}P_2(i+1)=\\
&\delta_{id}\cdots\delta_{(i-1)d+1}\left[p^1_2((i-1)d+1)\cdots p^d_2((i-1)d+1)\right]^{-1}P_2(i)\left(\Delta(i)\right)^{-1}\\
\end{split}
\end{equation}

At this point, we emphasize that their exist a deterministic matrix valued function $\underline{\Delta}$ such that for all $i\in\N^*$, $\Delta_i=\underline{\Delta}\left(\zeta^{d(i-1)+1},\dots,\zeta^{di}\right)$. In the same way, we define $\underline{P}_1$ and $\underline{P}_2$. The RHS of (\ref{eq1}) is a function of $\zeta^{(i-1)d+1},\dots,\zeta^{id}$ whereas the LHS is a matrix valued function of $\zeta^{id+1},\dots,\zeta^{d(i+1)}$, thus, both functions are constant. Therefore, there exist a matrix ${\cal I}$ such that for all $i\in\N^*$
\begin{equation}
\label{constant}
P_2(i+1)=p^1_2(id+1)\cdots p^d_2((id+1)){\cal I}.
\end{equation}
Therefore
\[\Delta_i={\cal I}^{-1}\delta_{id}\cdots\delta_{(i-1)d+1}{\cal I}.\]

Note that (\ref{constant}) can be rephrased in the following way. $\underline{P}_2$ and $\underline{p}^1_2\cdots\underline{p}^d_2$ are equal up to multiplication by a constant to ${\cal I}$. Therefore, to prove that ${\cal I}=\Id_{2d}$ for the choice for $\underline{p}^s_1$, $\underline{p}^s_2$ and $\underline{\delta}^s$ that we have made in Lemma \ref{decomposition}, it is enough to prove that for one given value of $\zeta^{1},\dots,\zeta^{d}$,
\[\underline{P}_2\left(\zeta^1,\dots,\zeta^d\right)=\underline{p}^1_2(\zeta^d)\cdots\underline{p}^d_2(\zeta^1).\]
We will prove it for $\zeta^1=\cdots=\zeta^d=(1,0,\dots,0,1)$. Indeed,
\[\underline{P}_2((1,0,\dots,0,1),\dots,(1,0,\dots,0,1))=\begin{pmatrix}
0_d&\Id_d\\
\Id_d&-\Id_d
\end{pmatrix}.\]
For $1\leq s\leq d$,
\[\underline{p}^s_2((1,0,\dots,0,1))=\left(\begin{array}{c|c}
 & \\
0_{2d-1,1}&\Id_{2d-1}\\
 & \\\hline
1&
\begin{array}{@{\hspace{1cm}}c@{\hspace{1cm}}|c|@{\hspace{1cm}}c@{\hspace{1cm}}}
0_{1,d-1}&-1&0_{1,d-1}
\end{array}
\end{array}\right),\]
Hence, by induction on $1\leq t\leq d$, $\underline{p}^1_2((1,0,\dots,0,1))\cdots\underline{p}^t_2((1,0,\dots,0,1))=$
\[\left(\begin{array}{c|c}
 & \\
0_{2d-t,t}&\Id_{2d-t}\\
 & \\\hline
\Id_t&
\begin{array}{@{\hspace{10pt}}c@{\hspace{10pt}}|c|@{\hspace{10pt}}c@{\hspace{10pt}}}
0_{t,d-t}&-\Id_t&0_{t,d-t}
\end{array}
\end{array}\right).\]
Therefore, ${\cal I}=\Id_{2d}$.
\end{proof}

%%%%%%%%%%%%%%%%%%%%%%%%%%%%%%%%%%%%%%%%%
%%%%%%%%%%%%%%%%%%%%%%%%%%%%%%%%%%%%%%%%%
%%%%%%%%%%%%%%%%%%%%%%%%%%%%%%%%%%%%%%%%%

\subsubsection{Proof of Corollary \ref{1stepthouless}}
\label{sec:ProofOfCorollaryRef1stepthouless}

%%%%%%%%%%%%%%%%%%%%%%%%%%%%%%%%%%%%%%%%%
%%%%%%%%%%%%%%%%%%%%%%%%%%%%%%%%%%%%%%%%%
%%%%%%%%%%%%%%%%%%%%%%%%%%%%%%%%%%%%%%%%%

Let us compute $\E\log\nrm{\mathfrak{n}_{d+1}}$. To that extent, we define $(e_1,\dots,e_{2d})$ the canonical basis of $\C^{2d}$ and we take $(e_{i_1}\wedge\cdots\wedge e_{i_d}|1\leq i_1<\cdots<i_d\leq 2d)$ as a basis of $\C^{\binom{2d}{d}}$. For given $1\leq i_1<\cdots<i_d\leq 2d$ and $1\leq j_1<\cdots<j_d\leq 2d$ the coefficient of $\mathfrak{n}_{d+1}(e_{i_1}\wedge\cdots\wedge e_{i_d})$ in $e_{j_1}\wedge\cdots\wedge e_{j_d}$ (we denote by $a$ its absolute value) is the determinant of the $d\times d$ sub-matrix of $\mu_{d+1}$ obtained by taking the lines $1\leq j_1<\cdots<j_d\leq 2d$ and the columns $1\leq i_1<\cdots<i_d\leq 2d$; we denote the latter sub-matrix by ${\cal D}$. Denote by $\widetilde{\zeta}_{i,l}$ the coefficient at position $(i,l)$ of $G_{dn}$.
\begin{itemize}
	\item If $1\leq j_1<\cdots<j_d\leq 2d-1$,
	\begin{itemize}
		\item if for all $1\leq s\leq d$, $i_s=j_s+1$, then $a=1$;
		\item otherwise, there exists a line of zeros in ${\cal D}$, therefore, $a=0$.
	\end{itemize}
	\item If $1\leq j_1<\cdots<j_{d-1}\leq 2d-1$, and $j_d=2d$,
	\begin{itemize}
		\item if there exists $1\leq s_0\leq d-1$ such that for all $1\leq s<s_0$, $i_s=j_s+1$, for all $s_0< s\leq d$, $i_s=j_{s-1}+1$ and $j_{s_0}=l\not\in\{j_1,\dots,j_{d-1}\}$, then $a=\abs{\widetilde{\zeta}_{d+1,l}/\widetilde{\zeta}_{d+1,2d+1}}$;
		\item otherwise, there exists a line of zeros in ${\cal D}$, therefore, $a=0$.
	\end{itemize}
\end{itemize}

We now count how many times each value appears as the absolute value of a coefficient of $\mathfrak{n}_{d+1}$.
\begin{itemize}
	\item To pick $1$, one needs to pick $d$ lines among the first $2d-1$ lines of $\mu_{d+1}$ and then, one has no choice for the columns: $\binom{2d-1}{d}$ choices.
	\item To pick $\abs{\widetilde{\zeta}_{d+1,1}/\widetilde{\zeta}_{d+1,2d+1}}$, one needs to pick $d-1$ lines among the first $2d-1$ lines of $\mu_{d+1}$ and then, one has no choice for the remaining line and the columns: $\binom{2d-1}{d-1}=\binom{2d-1}{d}$ choices.
	\item To pick $\abs{\widetilde{\zeta}_{d+1,l}/\widetilde{\zeta}_{d+1,2d+1}}$ for a given $2\leq l\leq2d$, one needs to pick $d-1$ lines among the first $2d-1$ lines of $\mu_{d+1}$ and one cannot pick the $(k-1)$-th line. Then one has no choice for the remaining line and the columns: $\binom{2d-2}{d-1}$ choices.
\end{itemize}

We factorize the term $1/\abs{\widetilde{\zeta}_{d+1,2d+1}}$, whose log-expectation cancels out with $\E_{\pi_0,\pi_d}\log\abs{\zeta_0\zeta_d^\dagger}$ and get the claimed bound.

%%%%%%%%%%%%%%%%%%%%%%%%%%%%%%%%%%%%%%%%%
%%%%%%%%%%%%%%%%%%%%%%%%%%%%%%%%%%%%%%%%%
%%%%%%%%%%%%%%%%%%%%%%%%%%%%%%%%%%%%%%%%%

\subsubsection{Proof of Proposition \ref{d2}}
\label{sec:ProofOfPropositionRefD2}

%%%%%%%%%%%%%%%%%%%%%%%%%%%%%%%%%%%%%%%%%
%%%%%%%%%%%%%%%%%%%%%%%%%%%%%%%%%%%%%%%%%
%%%%%%%%%%%%%%%%%%%%%%%%%%%%%%%%%%%%%%%%%

According to Proposition \ref{highSNR},
\begin{align*}
\Linf&=\frac{-1}{\log2}\Big[\log\rho+\E\log\abs{\zeta_0\zeta_d^\dagger}+\\
&\qquad\qquad\max\left(\E\log\abs{\det\psi_1^1}\,;\,\E\log\abs{\det\psi_1^2}\,;\, \gamma(\psi^1)+\gamma(\psi^2)\right)\Big],
\end{align*}
where
\[\psi_i^1=\begin{pmatrix}-\frac{\beta b_i}{\alpha a_i}&1\\-\frac{c_i}{\alpha a_i}&0\end{pmatrix}\textrm{ and }
\psi_i^2=\begin{pmatrix}-\frac{\beta b_i^\dagger}{c_i^\dagger}&-\frac{\alpha a_i^\dagger}{c_i^\dagger}\\1&0\end{pmatrix}.\]
Therefore,
\begin{align*}
\Linf&=\frac{-1}{\log2}\max\big(2\E\log\abs{a_1}\,;\,2\E\log\abs{a_1}+2\log\alpha\,;\\ &\qquad\qquad\qquad\qquad2\E\log\abs{a_1}+\log\alpha+\gamma(\psi^1)+\gamma(\psi^2)\big).
\end{align*}
Since $\alpha\leq1$, $\log\alpha\leq0$, therefore
\begin{equation}
\label{max}
\Linf=\frac{-1}{\log2}\max\left(2\E\log\abs{a_1}\,;\,2\E\log\abs{a_1}+\gamma\left(\widetilde{\psi}^1\right)+\gamma(\psi^2)\right),
\end{equation}
where
\[\widetilde{\psi}_i^1=\alpha\psi_i^1=\begin{pmatrix}-\frac{\beta b_i}{a_i}&\alpha\\-\frac{c_i}{a_i}&0\end{pmatrix}.\]
In order to finish the proof, we will construct of family of functions $(f_p)_{p\in\N^*}$ from $(0,1]\times[0,1]$ to $\R$ such that for all $p\in\N^*$, $f_p(\alpha,\beta)$ is non-decreasing in $\alpha$ and in $\beta$ and such that for all $p\in\N^*$ and for all $(\alpha,\beta)\in(0,1]\times[0,1]$,
\[\gamma\left(\widetilde{\psi}^1\right)+\gamma(\psi^2)\leq f_p(\alpha,\beta).\]
We define ${\cal D}$ in the following way:
\[{\cal D}\triangleq\bigcup_{p\in\N^*}\{(\alpha,\beta)\in(0,1]\times[0,1]\ ;\ f_p(\alpha,\beta)\leq0\}.\]

Since for all $p\in\N^*$, $f_p(\alpha,\beta)$ is non-decreasing in $\alpha$ and in $\beta$, we get that for all $(x,y)\in{\cal D}$, $(0,x)\times[0,y)\subset{\cal D}$. Moreover, by (\ref{max}), if $(\alpha,\beta)\in{\cal D}$, then (\ref{highSNRd2}) is verified.

Fix $p\in\N^*$. First note that by (\ref{subadditive}),
\[\gamma\left(\widetilde{\psi}^1\right)+\gamma(\psi^2)\leq\frac{1}{p}\left(\E\log\nrm{\widetilde{\psi}^1_p\cdots \widetilde{\psi}^1_1}+\E\log\nrm{\psi^2_p\cdots \psi^2_1}\right).\]
Recall that we use the Fr\"obenius norm on matrices. Denote $\phi^1(\alpha,\beta)=\widetilde{\psi}^1_p\cdots \widetilde{\psi}^1_1$ and $\phi^2(\alpha,\beta)=\psi^2_p\cdots \psi^2_1$. Note that the coefficients of $\phi^1(\alpha,\beta)$ and $\phi^2(\alpha,\beta)$ are polynomials in $\alpha$ and $\beta$. The function $1/p\left(\E\log\nrm{\phi^1(\alpha,\beta)}+\E\log\nrm{\phi^2(\alpha,\beta)}\right)$ would be a good candidate for $f_p$ but it is not non-decreasing in $\alpha$ and $\beta$, therefore, we have to modify it slightly.

Consider $P$ a polynomial in $\alpha$ and $\beta$,
\[P(\alpha,\beta)=\sum_{i,j=1}^n \theta_{i,j}\alpha^i\beta^j.\]
Define the polynomial $\abs{P}$ in the following way
\[\abs{P}(\alpha,\beta)=\sum_{i,j=1}^n \abs{\theta_{i,j}}\alpha^i\beta^j.\]
By the triangle inequality, for all $(\alpha,\beta)\in(0,1]\times[0,1]$, $\abs{P(\alpha,\beta)}\leq\abs{P}(\alpha,\beta)$. Moreover, $\abs{P}(\alpha,\beta)$ is non decreasing for $(\alpha,\beta)\in(0,1]\times[0,1]$.

Define the matrices $\abs{\phi^1}(\alpha,\beta)$ and $\abs{\phi^2}(\alpha,\beta)$ in the following way.

\noindent For $i,j,k=1,2$, set $\abs{\phi^k}_{i,j}=\abs{\phi^k_{i,j}}$. Then,
\[\nrm{\phi^1(\alpha,\beta)}\leq\nrm{\,\abs{\phi^1}(\alpha,\beta)}\textrm{ and }\nrm{\phi^2(\alpha,\beta)}\leq\nrm{\,\abs{\phi^2}(\alpha,\beta)}.\]
Moreover $\nrm{\,\abs{\phi^1}(\alpha,\beta)}$ and $\nrm{\,\abs{\phi^2}(\alpha,\beta)}$ are non decreasing for $(\alpha,\beta)\in(0,1]\times[0,1]$. Thus, we conclude the proof by defining
\[f_p=\frac{1}{p}\left(\E\log\nrm{\,\abs{\phi^1}(\alpha,\beta)}+\E\log\nrm{\,\abs{\phi^2}(\alpha,\beta)}\right).\]

\begin{rmk}
Note that if we define
\[\abs{\widetilde{\psi}_i^1}=\begin{pmatrix}\frac{\beta \abs{b_i}}{\abs{a_i}}&\alpha\\\frac{\abs{c_i}}{\abs{a_i}}&0\end{pmatrix}\textrm{ and }
\abs{\psi_i^2}=\begin{pmatrix}\frac{\beta \abs{b_i}}{\abs{c_i}}&\frac{\alpha \abs{a_i}}{\abs{c_i}}\\1&0\end{pmatrix},\]
then
\[f_p=1/p\left(\E\log\nrm{\,\abs{\widetilde{\psi}^1_p}\cdots \abs{\widetilde{\psi}^1_1}\,}+\E\log\nrm{\,\abs{\psi^2_p}\cdots \abs{\psi^2_1}\,}\right).\]
We use that fact in the numerical computation of the functions $f_p$.
\end{rmk}

%%%%%%%%%%%%%%%%%%%%%%%%%%%%%%%%%%%%%%%%%
%%%%%%%%%%%%%%%%%%%%%%%%%%%%%%%%%%%%%%%%%
%%%%%%%%%%%%%%%%%%%%%%%%%%%%%%%%%%%%%%%%%

\subsection{Exterior product}
\label{sec:ExteriorProduct}

%%%%%%%%%%%%%%%%%%%%%%%%%%%%%%%%%%%%%%%%%
%%%%%%%%%%%%%%%%%%%%%%%%%%%%%%%%%%%%%%%%%
%%%%%%%%%%%%%%%%%%%%%%%%%%%%%%%%%%%%%%%%%

In this section we give the material on exterior products. We provide only the properties relevant to the article, see \cite[Chapter XVI.6-7]{algebra} and \cite[Chapter A.III.5]{bougerol-lacroix} for more details.

\begin{prop}
For $0\leq k\leq p$, the exterior product of $k$ vectors in $\C^p$, $v_1,\dots,v_k$ is denoted by $v_1\wedge\cdots\wedge v_k$. Is is a vector of the exterior product of degree $k$ of $\C^p$ that we denote by $\bigwedge^k\C^p$. $\bigwedge^k\C^p$ is a $\C$-vector space of dimension $\binom{k}{p}$.

The exterior product $v_1,\dots,v_k$ is a multi-linear (i.e. linear in every $v_i$, $1\leq i\leq k$) and anti-symmetric (i.e. $v_{\sigma(1)}\wedge\cdots v_{\sigma(k)}=\ep(\sigma)$ for $\sigma$ permutation of $\{1,\dots,k\}$ and $\ep(\sigma)$ its signature) function.

If $e_1,\dots,e_p$ is a basis of $\C^p$, then $(e_{i_1}\wedge\cdots\wedge e_{i_k}|1\leq i_1<\cdots<i_k\leq p)$ is a basis of $\bigwedge^k\C^p$. The later is called the \emph{canonical basis} of $\bigwedge^k\C^p$ if $e_1,\dots,e_p$ is the canonical basis of $\C^p$.

If $M$ is a matrix of size $p\times q$, the exterior product of $M$ that we denote by $\bigwedge^k M$ is a map from $\bigwedge^k\C^q$ to $\bigwedge^k\C^p$ such that
\[\bigwedge^k M\left(v_1\wedge\cdots\wedge v_k\right)=Mv_1\wedge\cdots\wedge Mv_k.\]

Finally, for two matrices $M$ and $N$, $\bigwedge^k\left(MN\right)=\bigwedge^k\left(M\right)\bigwedge^k\left(N\right)$.
\end{prop}

\begin{prop}
If $X$ is a square matrix of size $p$, then
\[\bigwedge^p X = \det X.\]
Moreover
\[\det \bigwedge^p X = \left(\det X\right)^p\]
\end{prop}

%%%%%%%%%%%%%%%%%%%%%%%%%%%%%%%%%%%%%%%%%
%%%%%%%%%%%%%%%%%%%%%%%%%%%%%%%%%%%%%%%%%
%%%%%%%%%%%%%%%%%%%%%%%%%%%%%%%%%%%%%%%%%

\subsection{Capacity of the non-fading channels}
\label{sec:CapacityOfTheNonFadingChannels}

%%%%%%%%%%%%%%%%%%%%%%%%%%%%%%%%%%%%%%%%%
%%%%%%%%%%%%%%%%%%%%%%%%%%%%%%%%%%%%%%%%%
%%%%%%%%%%%%%%%%%%%%%%%%%%%%%%%%%%%%%%%%%

In this Section, we give expressions of the limiting sum-rate per-cell capacity for the Soft-Handoff model and the Wyner model (both symmetric and asymmetric) for the non-fading channels. Those expressions are consequences of results on Toeplitz matrices \cite{Gray-paper-72}. See \cite	{Wyner-94} for an example of derivation.

%%%%%%%%%%%%%%%%%%%%%%%%%%%%%%%%%%%%%%%%%
%%%%%%%%%%%%%%%%%%%%%%%%%%%%%%%%%%%%%%%%%
%%%%%%%%%%%%%%%%%%%%%%%%%%%%%%%%%%%%%%%%%

\subsubsection{The Soft-Handoff model}
\label{sec:TheSoftHandoffModel}

%%%%%%%%%%%%%%%%%%%%%%%%%%%%%%%%%%%%%%%%%
%%%%%%%%%%%%%%%%%%%%%%%%%%%%%%%%%%%%%%%%%
%%%%%%%%%%%%%%%%%%%%%%%%%%%%%%%%%%%%%%%%%

We assume that $d=1$, and for $i\in\N^*$, $\zeta_{i,i+1}=\alpha\in[0,1]$ and $\zeta_{i,i}=1$. Then, the limiting per-cell sum-rate capacity is
\[Cap(\rho)=\log\left(\frac{1+K\rho\left(1+\alpha^2\right)+\sqrt{1+2K\rho\left(1+\alpha^2\right)+K^2\rho^2\left(1-\alpha^2\right)^2}}{2}\right).\]

%%%%%%%%%%%%%%%%%%%%%%%%%%%%%%%%%%%%%%%%%
%%%%%%%%%%%%%%%%%%%%%%%%%%%%%%%%%%%%%%%%%
%%%%%%%%%%%%%%%%%%%%%%%%%%%%%%%%%%%%%%%%%

\subsubsection{The Wyner model}
\label{sec:TheWynerModel}

%%%%%%%%%%%%%%%%%%%%%%%%%%%%%%%%%%%%%%%%%
%%%%%%%%%%%%%%%%%%%%%%%%%%%%%%%%%%%%%%%%%
%%%%%%%%%%%%%%%%%%%%%%%%%%%%%%%%%%%%%%%%%

%%%%%%%%%%%%%%%%%%%%%%%%%%%%%%%%%%%%%%%%%
%%%%%%%%%%%%%%%%%%%%%%%%%%%%%%%%%%%%%%%%%
%%%%%%%%%%%%%%%%%%%%%%%%%%%%%%%%%%%%%%%%%

\paragraph*{The symmetric setting}
\label{sec:TheSymmetricSetting}
%%%%%%%%%%%%%%%%%%%%%%%%%%%%%%%%%%%%%%%%%
%%%%%%%%%%%%%%%%%%%%%%%%%%%%%%%%%%%%%%%%%
%%%%%%%%%%%%%%%%%%%%%%%%%%%%%%%%%%%%%%%%%

We assume that $d=2$, and for $i\in\N^*$, $\zeta_{i,i+2}=\zeta_{i,i}=\alpha\in[0,1]$ and $\zeta_{i,i+1}=1$. Then, the limiting per-cell sum-rate capacity is
\[Cap(\rho)=\int_0^1\log\left(1+K\rho\left(1+2\alpha\cos(2\pi f)\right)^2\right)df.\]

%%%%%%%%%%%%%%%%%%%%%%%%%%%%%%%%%%%%%%%%%
%%%%%%%%%%%%%%%%%%%%%%%%%%%%%%%%%%%%%%%%%
%%%%%%%%%%%%%%%%%%%%%%%%%%%%%%%%%%%%%%%%%

\paragraph*{The asymmetric setting}
\label{sec:TheAsymmetricSetting}

%%%%%%%%%%%%%%%%%%%%%%%%%%%%%%%%%%%%%%%%%
%%%%%%%%%%%%%%%%%%%%%%%%%%%%%%%%%%%%%%%%%
%%%%%%%%%%%%%%%%%%%%%%%%%%%%%%%%%%%%%%%%%

We assume that $d=2$, and for $i\in\N^*$, $\zeta_{i,i+2}=\zeta_{i,i+1}=\alpha\in[0,1]$ and $\zeta_{i,i}=1$. Then, the limiting per-cell sum-rate capacity is
\[Cap(\rho)=\int_0^1\log\left(1+K\rho\left(1+2\alpha^2+2\alpha(1+\alpha)\cos(2\pi f)+2\alpha\cos(4\pi f)\right)\right)df.\]

\bibliographystyle{ieeetr}
\bibliography{Reference_List}

\end{document}